\documentclass[%
reprint,
amsmath,amssymb,
aps,prd,
showkeys
]{revtex4-2}

\usepackage{longtable}
\usepackage{graphicx}
\usepackage{dcolumn}
\usepackage{bm}
\usepackage{hyperref}
\hypersetup{
    colorlinks,
    citecolor=red,
    filecolor=black,
    linkcolor=blue,
    urlcolor=black
}
\usepackage{nicematrix}
\usepackage{tikz}
\usepackage{orcidlink}
\usepackage[ruled, noline, noend]{algorithm2e}
\usepackage{mathtools}

\usepackage[arrowdel]{physics}
\newcommand{\eps}{\varepsilon}

\DeclareMathOperator*{\argmax}{argmax}

\begin{document}

\title{Quantum error mitigation by hierarchy-informed sampling:\\chiral dynamics in the Schwinger model}
\date{\today}

\author{Theo Saporiti \orcidlink{0009-0008-9738-3402}}
\email{theo.saporiti@cea.fr}
\author{Oleg Kaikov \orcidlink{0000-0002-9473-7294}}
\author{Vasily Sazonov \orcidlink{0000-0002-8152-0221}}
\author{Mohamed Tamaazousti \orcidlink{0000-0002-3947-9069}}
\affiliation{Université Paris-Saclay, CEA, List, F-91120, Palaiseau, France}

\begin{abstract}\noindent
Quantum simulations on current NISQ hardware are limited by its noisy nature, making efficient quantum error mitigation methods highly demanded. In this paper we introduce a novel mitigation scheme, applicable to arbitrary quantum simulations of time-dependent Hamiltonian dynamics on NISQ devices. The scheme uses a polynomial subset of extended qubit Bogoliubov-Born-Green-Kirkwood-Yvon (BBGKY) hierarchy equations as a sampling criterion of possible mitigated candidates for the quantum observables. We show that for favorable Hamiltonians the polynomial subset of BBGKY hierarchy equations leads to a polynomial overhead in both classical and quantum resources. We employ the method to mitigate simulations of the chiral magnetic effect (CME), a chiral feature of the Schwinger model. We empirically show the effectiveness of our scheme at recovering the real-time dynamics of the CME from noisy quantum simulations of the Schwinger model, for a range of different parameter values of the model. We numerically demonstrate a systematic reduction of quantum noise, together with an increasing noise reduction capability as the amount of BBGKY constraints grows.
\end{abstract}


\maketitle

\section{Introduction}

The future advent of noiseless quantum computing promises quantum advantage, notably in the simulation of the real-time dynamics of many-body physical systems. Indeed, quantum simulations allow for calculations in regimes that are currently untractable by classical hardware \cite{Banuls:2019rao, Banuls:2019bmf, Funcke:2023jbq, DiMeglio:2023nsa}. For instance, the high-density regions of the quantum chromodynamics (QCD) phase diagram cannot be studied with conventional classical Monte Carlo methods: the presence of a non-zero baryon chemical potential causes the sign problem \cite{deForcrand:2009zkb, Nagata:2021ugx}, of which crucially quantum simulations are naturally free \cite{Funcke:2023jbq, DiMeglio:2023nsa}. However, the concrete application of present quantum devices is spoiled by the systematic occurrence of quantum noise within simulations \cite{Steane:1997kb, DiVincenzo:2000tra, Ladd:2010cup, Georgescu:2013oza, Preskill:2018jim}. While fault-tolerant protocols do exist \cite{ShorEC, Steane, Calderbank, Nielsen_Chuang_2010}, their use is obstructed by the small number of qubits in current noisy intermediate-scale quantum (NISQ) computers \cite{Chatterjee:2024kpt}. It is in this context that less resource-intensive quantum error mitigation techniques were developed \cite{PhysRevApplied.20.064027, PhysRevA.94.052325, PhysRevA.105.032620, Temme, Li:2016vmf, Giurgica-Tiron:2020rcf, PracticalQEM, LearningBasedQEM, AIQEM, Cai:2020khs, Saporiti:2025zwf, Kaikov:2025dtr}.

In this paper, we introduce and empirically investigate the effectiveness of a novel mitigation technique, designed to mitigate the noisy simulations of arbitrary time-dependent Hamiltonians. The general idea behind our scheme is that physical knowledge of ideal noiseless dynamics can be used to reduce the error in realistic noisy quantum measurements (see for example \cite{Bonet-Monroig:2018mpi, Smart:2019hyn}).

In our previous work \cite{Saporiti:2025zwf}, we explored such an approach by employing Bogoliubov-Born-Green-Kirkwood-Yvon (BBGKY) \cite{Bogoliubov, Born:1946vqa, kirkwood, yvon1935theorie} equations as supplementary constraints to the specifically chosen zero-noise extrapolation (ZNE) procedure \cite{Temme, Li:2016vmf}, restricted to real-time dynamics for time-independent Hamiltonians with one-site and two-site interactions. It was numerically found that our previous BBGKY-improved ZNE method was able to systematically reduce quantum error \cite{Saporiti:2025zwf}. However, the method could not guarantee an improvement of its error mitigation capability as the number of selected BBGKY equations increased. Nonetheless, our earlier work demonstrated the potential utility of the BBGKY hierarchy in the context of quantum mitigation.

In this paper, we propose a novel mitigation scheme which is fully independent from \cite{Saporiti:2025zwf}, that is, our method is a new standalone mitigation technique independent of ZNE or any other preexisting schemes. Contrarily to \cite{Saporiti:2025zwf}, our method is applicable to Hamiltonians which can be time-dependent and which can encode interactions among an arbitrary number of spins. To this end, we formulate an extended qubit BBGKY hierarchy, associated to any such quantum simulation, and we investigate its properties, allowing us to gain an improved understanding of its underlying structure. Our scheme employs a polynomially-large portion of the BBGKY hierarchy, used as a sampling criterion of possible quantum mitigations. These are selected in a Markov Chain Monte Carlo \cite{Metropolis:1953am, Hastings:1970aa} simulated annealing \cite{Kirkpatrick:1983zz} procedure, according to their physicality towards the chosen portion of constraints from the hierarchy. As the candidate mitigations encode portions of the ideal noiseless dynamics, their average provides a physics-informed quantum mitigation of the original noisy measurements, thus a quantum error reduction capability is obtained.

The method is empirically put to the test on noisy quantum simulations of the Schwinger model \cite{PhysRev.128.2425}. We empirically demonstrate the effectiveness of our method at recovering the real-time dynamics of the chiral magnetic effect (CME) \cite{Fukushima:2008xe, Fukushima:2010vw, Kharzeev:2013ffa, Kharzeev:2023zbo, Kharzeev:2024zzm, Warringa:2008kv}, the generation of an electric current due to a chiral imbalance in a strong magnetic field, and we numerically show that its recovery capability improves as the number of BBGKY constraints increases.

The Schwinger model bears resemblances with key properties of QCD, such as the sign problem \cite{Gattringer:2015nea}, color-confinement \cite{Coleman:1975pw} and the CME \cite{PhysRevResearch.2.023342}, while benefiting from a relative simplicity compared to QCD. As a result, its quantum spin chain realization is commonly used as a benchmark toy model to demonstrate the future applicability of current methods to QCD simulations \cite{PhysRevD.105.014504,Pederiva:2022br,schwinger,Yamamoto:2022Qn,PhysRevD.105.094503,Ghim:2024pxe,DAnna:2024mmz,Kaikov:2024acw,Saporiti:2025zwf,Kaikov:2025dtr}. We apply our mitigation scheme specifically to simulations of the CME for a variety of reasons, which we now list.
First, the CME bears importance in relativistic heavy-ion collisions \cite{Kharzeev:2013ffa, Kharzeev:2023zbo, Kharzeev:2024zzm}: indeed, its macroscopic observation would imply the existence of topological transitions in the produced quark-gluon plasma \cite{Kharzeev:2023zbo}, providing a possible explanation to the matter-antimatter asymmetry problem \cite{Sakharov:1967dj}. Then, the CME represents one of the key obstacles in QCD computations: the plasma expansion after a heavy-ion collision event is a dynamical process happening in dense quark matter \cite{Kharzeev:2024zzm}, leading to the sign problem. In terms of practical applications, the CME as observed within Dirac \cite{doi:10.1126/science.aac6089} and Weyl \cite{PhysRevX.5.031023} semimetals can be used to control corresponding chiral qubits \cite{Kharzeev:2019ceh}: a better understanding of the CME inside these materials would directly contribute to the development of quantum hardware. Lastly, the CME in the context of the Schwinger model also provides a time-dependent phenomenon, which is the target application of our method.

This paper is organized as follows. In Sec.~\ref{sec:bbgky} we derive the extended qubit BBGKY equations (Sec.~\ref{sec:generalized_bbgky}) and discuss their hierarchical arrangement (Sec.~\ref{sec:hierarchical_structure}). In Sec.~\ref{sec:mitigation}, after a reminder on Trotterization (Sec.~\ref{sec:trotterization}), we justify (Sec.~\ref{sec:probabilistic_solutions}) and construct (Sec.~\ref{sec:our_scheme}) our mitigation scheme, based on the random sampling of possible mitigations (Sec.~\ref{sec:metropolis}). In Sec.~\ref{sec:schwinger} we briefly describe the CME in QCD (Sec.~\ref{sec:cme_qcd}) and in the Schwinger model (Sec.~\ref{sec:cme}). In Sec.~\ref{sec:results} we empirically assess the effectiveness of the mitigation method at recovering the CME dynamics from noisy quantum simulations (Sec.~\ref{sec:simulations} and \ref{sec:numerical_results}, framework and results respectively). We also verify the general theoretical properties of the BBGKY hierarchy in the case of the Schwinger model (Sec.~\ref{sec:hierarchy_visualization}). Finally in Sec.~\ref{eq:conclusions} we draw conclusions and propose future extensions of this work.

\section{Extended qubit BBGKY hierarchy}\label{sec:bbgky}

In this Section, we derive and discuss the properties of the extended qubit BBGKY hierarchy, which encodes the real-time dynamics of arbitrary time-dependent quantum (long-range interacting) spin-$\frac{1}{2}$ chain systems. We discuss the prerequisites for the amount of observables in the selected BBGKY equations to scale polynomially in the number of qubits. This suitable feature allows the application of the BBGKY hierarchy for quantum computing purposes, in particular for quantum error mitigation. Here, and throughout the work, we set $\hbar = c = 1$. Moreover, a complete listing of all the symbols employed in this paper can be found in Appendix \ref{app:symbols}.

\subsection{Hamiltonians and Pauli strings}

Consider $N_\text{Q}$ qubits each uniquely labeled by an integer $i \in S := \qty{1, \dots, N_\text{Q}}$, called site. Then, $S$ represents the whole system, $\abs{S} = N_\text{Q}$, and $A \in \mathcal{P}(S)$ represents a subsystem, where $\mathcal{P}(S)$ is the power set of $S$
\begin{equation}
    \mathcal{P}(S) := \qty{\qty{1}, \dots, \qty{N_\text{Q}}, \qty{1,2}, \dots, S},
\end{equation}
which we define without the empty set $\varnothing \not\in \mathcal{P}(S)$. The interactions between all qubits, generating the full dynamics of the (long-range interacting) spin chain across the time window $t \in [0, T]$, are encoded in the time-dependent Hamiltonian
\begin{equation}\label{eq:hamiltonian_general}
    H(t) := \sum_{(B,b) \in \mathcal{H}} h^b_B(t) \sigma^b_B.
\end{equation}
In the above, $h^b_B(t)$ is the coupling at time $t$ associated with Pauli string $\sigma_B^b$, the latter being defined below and encoded in the tuple $(B,b)$, and $\mathcal{H}$ denotes the set of all (encoded) interaction terms
\begin{equation}
\begin{split}
    \mathcal{H} := &\{(B,b) \in \mathcal{P}(S) \times \mathcal{R}(S) \mid\\
    &\;\;1 \leq \abs{B} = \abs{b} \quad\text{and}\quad \exists t \in [0, T] \colon h^b_B(t) \neq 0\},
\end{split}
\end{equation}
where $\mathcal{R}(S)$ is the equivalent of $\mathcal{P}(S)$ for the directions of the Pauli operators
\begin{equation}
\begin{split}
    \mathcal{R}(S) &:= \bigcup_{n=1}^{\abs{S}} \qty{1,2,3}^n\\
    &= \qty{(1), (2), (3), (1,1), (1,2), \dots, (3, \dots, 3)},
\end{split}
\end{equation}
with the important caveat that $\mathcal{R}(S)$ is not the set of all subsets $A \in \mathcal{P}(S)$ of $S$, but the set of all (ordered) $n$-tuples $a \in \mathcal{R}(S)$ of elements $\qty{1,2,3}$, where $1 \leq n \leq \abs{S}$ \footnote{As an example, consider $A = \qty{1,2} = \qty{2,1}$, $a=(2,3)$ and $a'=(3,2) \neq a$. Then $(A, a)$ encodes Pauli string $\sigma^2_1 \sigma^3_2$, whereas $(A, a')$ encodes Pauli string $\sigma^3_1 \sigma^2_2$.}. Indeed, provided the number of components $\abs{a}$ in $a$ matches the number of elements of $A$, $1 \leq \abs{A} = \abs{a}$, any tuple $(A, a)$ defines a Pauli string
\begin{equation}\label{eq:pauli_string}
    \sigma^a_{A} := \prod_{i \in A} \sigma^{a_i}_i,
\end{equation}
where $\sigma^{a_i}_i$ is the Pauli operator of direction $a_i \in \qty{1,2,3}$ acting on site $i \in S$. Importantly, $a_i$ does not denote the $i$th component of $a$, but rather the direction associated to the $i$th site. More formally, the index of $a_i$ in $a$ is the same as the index of $i$ in $A$, if the elements of $A$ were sorted in ascending order. For the remainder of the paper, by abuse of notation we will refer to $\sigma^a_A$ as the encoded tuple $(A,a)$, and vice versa. Moreover, we will always consider $\abs{A}, \abs{B} \geq 1$.

\subsection{Extended qubit BBGKY equations}\label{sec:generalized_bbgky}

From now on, for notational convenience, we will omit to denote all explicit time dependencies.

Injecting \eqref{eq:hamiltonian_general} and \eqref{eq:pauli_string} into Ehrenfest's theorem \cite{Ehrenfest:1927swx}, as outlined in Appendix \ref{app:commutator}, gives
\begin{equation}\label{eq:bbkgy_equation}
\begin{split}
    &\dv{t} \ev{\sigma^a_A} = \sum_{(B,b) \in \mathcal{H}} h^b_B f^{ab}_{AB}\\
    &\qquad\cdot\ev{\prod_{i \in A \setminus B} \sigma^{a_i}_i \prod_{j \in B \setminus A} \sigma^{b_j}_j \prod_{k \in A \cap B} \qty(\delta_{a_k b_k} + \eps_{a_k b_k c} \sigma_k^c)},
\end{split}
\end{equation}
where $\delta$ is the 2-dimensional Kronecker delta, $\eps$ is the 3-dimensional Levi-Civita symbol, Einstein's summation convention applies to $c \in \qty{1,2,3}$, and where we define the factor
\begin{equation}\label{eq:f_factor_def}
    f_{AB}^{ab} := 2 (-1)^{(d_{AB}^{ab} - 1)/2} (d_{AB}^{ab} \text{ mod } 2),
\end{equation}
with
\begin{equation}\label{eq:differences_number}
    d_{AB}^{ab} := \sum_{k \in A \cap B} (1 - \delta_{a_k b_k})
\end{equation}
the amount of different directions among all sites in $A \cap B$. We call \eqref{eq:bbkgy_equation} the BBGKY equation associated to $\sigma^a_A$ and, for reasons that we will explain in the upcoming Sec.~\ref{sec:hierarchical_structure}, we call the set of all $4^{\abs{S}}$ possible BBGKY equations the BBGKY hierarchy of the system.

Every interaction term $(B,b) \in \mathcal{H}$ provides a corresponding correlator on the right-hand side (RHS) sum in \eqref{eq:bbkgy_equation}. We bound from above the total amount of such correlators $\abs{\mathcal{H}}$ appearing in any BBGKY equation as
\begin{equation}\label{eq:bound_number_RHS}
    \abs{\mathcal{H}} \leq \sum_{l=1}^{L_\mathcal{H}} 3^l \binom{N_\text{Q}}{l} \leq 3^{L_\mathcal{H} + 1} L_\mathcal{H} N_\text{Q}^{L_\mathcal{H}}.
\end{equation}
The first inequality is obtained by counting all possible Pauli strings of length $1 \leq l \leq L_\mathcal{H}$ up to the maximal length $L_\mathcal{H} := \max_{(B,b)\in\mathcal{H}}\abs{B} \leq N_\text{Q}$ of the longest Pauli string in $H$. The second inequality is obtained with the application of the known binomial coefficient upper bound $\binom{x}{y} \leq (ex/y)^y \leq 3x^y$, valid for any $0 < y \leq x$ \cite{knuth1968art}. In general, without knowing the dependence $L_\mathcal{H} = L_\mathcal{H}(N_\text{Q})$, the upper bound \eqref{eq:bound_number_RHS} cannot guarantee that $\abs{\mathcal{H}} \sim \text{poly}(N_\text{Q})$. However, depending on the definition of $\mathcal{H}$, it can still be $\abs{\mathcal{H}} \sim \text{poly}(N_\text{Q})$ even when the upper bound \eqref{eq:bound_number_RHS} is not polynomial in $N_\text{Q}$. This is for instance the case of the Schwinger model realization introduced in Sec.~\ref{sec:cme}, where although $L_\mathcal{H} = N_\text{Q}$ it is still $\abs{\mathcal{H}} \sim \text{poly}(N_\text{Q})$. Moreover, the polynomial scaling of $\abs{\mathcal{H}}$ obviously holds for a constant $L_\mathcal{H}$, such as in the Hamiltonian previously studied in \cite{Saporiti:2025zwf}. Indeed, by setting $L_\mathcal{H} = 2$ in \eqref{eq:bound_number_RHS}, one recovers (up to constant coefficients) the quadratic in $N_\text{Q}$ upper bound for the number of RHS correlators in any BBGKY equation found in \cite{Saporiti:2025zwf}.

\subsection{Hierarchical structure}\label{sec:hierarchical_structure}

\begin{figure}
\centering
\includegraphics[width=0.9\linewidth]{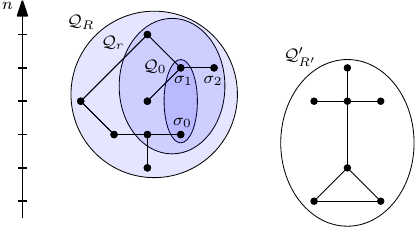}
\caption{Example of a BBGKY hierarchy and its structure. Each node represents a Pauli string of length $n$, while each edge represents an immediate connection. The full hierarchy can split into multiple disconnected subhierarchies, in this example $\mathcal{Q}_R$ and $\mathcal{Q}'_{R'}$. The sequence of $Q_r$ subsets, from $Q_0$ to the full subhierarchy $Q_R$, showcases the growth of the radius $r \in \qty{0,\dots,R}$. For instance, Pauli strings $\sigma_0$ and $\sigma_1$ are connected, while Pauli strings $\sigma_1$ and $\sigma_2$ are immediately connected.}
\label{fig:extended_hierarchy}
\end{figure}

The exponentially large set of all BBGKY equations encodes the full real-time dynamics of the spin chain, as the set of all Pauli strings forms an operator basis for any observable of the system. This set exhibits a hierarchical structure in the connections between different correlators, which we now explain.

We say $\sigma^c_C$ is downstream-connected to $\sigma^a_A$ and $\sigma^a_A$ is upstream-connected to $\sigma^c_C$, denoted as $\sigma^a_A \to \sigma^c_C$, if there exists a time $t$ for which $\sigma^c_C$ appears on the RHS of the BBGKY equation associated to $\sigma^a_A$. In Appendix \ref{app:symmetric_hierarchy} we show that $\sigma^a_A \to \sigma^c_C$ implies $\sigma^c_C \to \sigma^a_A$, hence we define the new symbol
\begin{equation}
    (\sigma^a_A \leftrightarrow \sigma^c_C) \quad \xLeftrightarrow{\text{def}} \quad (\sigma^a_A \to \sigma^c_C) \quad \text{or} \quad (\sigma^c_C \to \sigma^a_A)
\end{equation}
and say that $\sigma^a_A \leftrightarrow \sigma^c_C$ is an immediate connection between $\sigma^a_A$ and $\sigma^c_C$, or equivalently that $\sigma^a_A$ immediately connects to $\sigma^c_C$ (and vice versa). These immediate connections form an undirected graph, whose nodes represent all possible Pauli strings. By ordering the nodes with respect to their associated Pauli string length, this undirected graph becomes a hierarchy, in that \eqref{eq:bbkgy_equation} upstream-connects a generic $\sigma^a_A$ only to correlators $\sigma^c_C$ of specific length $\abs{C} = L^{ab}_{AB}$, where $(B,b) \in \mathcal{H}$. A concrete realization of such graph can be found in Fig.~\ref{fig:hierarchy} of the upcoming Sec.~\ref{sec:hierarchy_visualization}, which displays the BBGKY hierarchy associated to the Schwinger model investigated in this paper. While a simple expression for $L^{ab}_{AB}$ is difficult to obtain, in Appendix \ref{app:bounds} we derive the tightest upper and lower bounds
\begin{equation}\label{eq:length_bounds}
    \abs{\abs{A} - \abs{B}} + 1 \leq L^{ab}_{AB} \leq \abs{A} + \abs{B} - 1.
\end{equation}
As a limiting case of the above, by plugging in $\abs{B} \in \qty{1,2}$, one recovers the hierarchical structure previously studied in \cite{Cox:2018zxg, Saporiti:2025zwf}, namely $n$-point correlators being only immediately connected to $(n-1)$-point, $n$-point, and $(n+1)$-point correlators.

Any BBGKY hierarchy can potentially split into disconnected subhierarchies of independent correlators, that is, whose dynamics don't have a mutual influence across different subhierarchies. To see that, consider the sets of quantities $\mathcal{Q}_r \subseteq \mathcal{P}(S) \times \mathcal{R}(S)$ that are connected to an initial set of Pauli strings $\mathcal{Q}_0 \neq \varnothing$ by at most $r$ successive immediate connections. More precisely, we say that two Pauli strings $\sigma_A^a$ and $\sigma_C^c$ are connected (through the hierarchy) if $\sigma_A^a \leftrightarrow \dots \leftrightarrow \sigma_C^c$. Clearly $\mathcal{Q}_r \subseteq \mathcal{Q}_{r+1}$, and the set inclusion becomes a set equality for any radius $r$ bigger than or equal to the subhierarchy radius $R := \min(\argmax_{r \in \mathbb{N}_0} \abs{\mathcal{Q}_r})$, which defines the subhierarchy size $\abs{\mathcal{Q}_R}$. Overall, for $0 < r < r+1 < R$,
\begin{equation}\label{eq:inequalities}
    0 < \abs{\mathcal{Q}_0} < \dots < \abs{\mathcal{Q}_r} < \abs{\mathcal{Q}_{r+1}} < \dots < \abs{\mathcal{Q}_R}.
\end{equation}

Given a choice of $\mathcal{Q}_0$ and $r$, $\abs{\mathcal{Q}_{r+1}}$ is the amount of expectation values that have to be known in order to evaluate the $\abs{\mathcal{Q}_{r}}$ BBGKY equations associated to the elements of $\mathcal{Q}_r$. Importantly, if $\abs{\mathcal{H}} \sim \text{poly}(N_\text{Q})$, then by \eqref{eq:bbkgy_equation} the number of immediately connected correlators to any element of $\mathcal{Q}_r$ is $\text{poly}(N_\text{Q})$ as well. Hence, if $\abs{\mathcal{Q}_0} \sim \text{poly}(N_\text{Q})$, the amount of classical computational resources required to obtain $\mathcal{Q}_r$ for any fixed $r$ is polynomial in $N_\text{Q}$. This polynomial scaling in $N_\text{Q}$ renders the BBGKY hierarchy a suitable tool for quantum error mitigation purposes. Figure \ref{fig:extended_hierarchy} summarizes the concepts introduced in this Section.

\section{Mitigation technique}\label{sec:mitigation}

In this Section we propose our novel mitigation technique, based on a BBGKY physics-informed sampling of possible mitigation candidates. We begin with a reminder on Trotterization, to both introduce notations and to formally state the mitigation goal. We then explain and build the mathematical justification behind our scheme, namely the possibility to compute exact real-time dynamics as classical expectation values, stemming from a probability distribution encoding the BBGKY hierarchy. Next, we build over this framework our novel mitigation technique. By generalizing the probability distribution to account for the noisy quantum simulation measurements, the once exact probabilistic method becomes a mitigation procedure, informed by supplemented BBGKY constraints. Finally, we explain how to numerically implement the technique as a simulated annealing algorithm.

\subsection{Trotterized quantum simulations}\label{sec:trotterization}

Consider the real-time evolution of a given initial quantum state, from time $t=0$ to time $T$, obtained by a Trotterization involving $N_\text{T}$ Trotter slices of duration $\Delta t := T/N_\text{T}$, executed at time points $t_s := s \Delta t$ with $s \in \qty{0, \dots, N_\text{T}}$ \cite{82edc856-4d85-3b98-9b0d-ad55bb9315f6, Magnus:1954zz, Hatano:2005gh, Berry:2005yrf}. That is, we approximate the time-evolution operator as
\begin{equation}\label{eq:time_evolution_operator}
\begin{split}
    &\mathcal{T} \exp(-i\int_0^T\dd{t} H(t))\\
    &=\mathcal{T} \exp(-i\Delta t \sum_{s=1}^{N_\text{T}} \sum_{(B,b) \in \mathcal{H}} h_B^b (t_s) \sigma_B^b + \order{\Delta t})\\
    &= \mathcal{T}\prod_{s=1}^{N_\text{T}} \qty(\prod_{(B,b) \in \mathcal{H}} e^{-i (\Delta t) h^b_B(t_s) \sigma^b_B}) + \order{\Delta t},
\end{split}
\end{equation}
where in the first equality we approximate the integration of \eqref{eq:hamiltonian_general} with the rectangle quadrature formula, in the second equality we apply the Lie-Trotter formula \cite{82edc856-4d85-3b98-9b0d-ad55bb9315f6}, and the final product is understood to be time-ordered, denoted by $\mathcal{T}$. The $\order{\Delta t}$ error over the exact dynamics, which we call Trotter error, includes both the numerical integration error and the Lie-Trotter error.

We are interested in obtaining approximations of (linear combinations of) all $\qty{\ev{\sigma^a_A}}_{(A,a) \in \mathcal{Q}_0}$, which we label as $\qty{\ev{\sigma_q}}_{q=1, \dots, \abs{\mathcal{Q}_{r}}}$. Ultimately, we want to mitigate the quantum noise affecting their measurements. As explained in Sec.~\ref{sec:hierarchical_structure}, an expectation value can only affect another one if and only if their quantum operators are connected through the BBGKY hierarchy. Therefore, in the following and for the rest of the paper, given a fixed $r$, we will measure the extended set of expectation values $\qty{\ev{\sigma_q}}_{q=1, \dots, \abs{\mathcal{Q}_{r+1}}}$, namely all $\abs{\mathcal{Q}_{r+1}}$ Pauli strings appearing in the $\abs{\mathcal{Q}_{r}}$ BBGKY equations associated to $\mathcal{Q}_r$.

We denote with $\bar{x}_{qs}$ the measurement of $\ev{\sigma_q (t_s)}$, obtained as the average of $N_\text{S}$ shots. Importantly, for demonstration purposes, we assume that the initial quantum state is a polynomially-large linear combination of the computational quantum basis states spanning its exponentially-large Hilbert space. As a result, all $\abs{\mathcal{Q}_{r+1}} \sim \text{poly}(N_\text{Q})$ initial $t=0$ measurements $\bar{x}_{q0}$ can be efficiently evaluated on a classical machine up to floating-point error. Therefore, unless otherwise stated, from now on we will always assume $s \geq 1$. Finally, the definition of $\bar{x}_{qs}$ implies that it discretely distributes across $\bar{x}_{qs} \in \qty{-1 + k \Delta x : k \in \qty{0, \dots, N_\text{S}}} \subseteq [-1, 1]$, with spacing $\Delta x := 2/N_\text{S}$.

\subsection{Probabilistic solutions of the BBGKY hierarchy}\label{sec:probabilistic_solutions}

\begin{figure}
\centering
\includegraphics[width=0.65\linewidth]{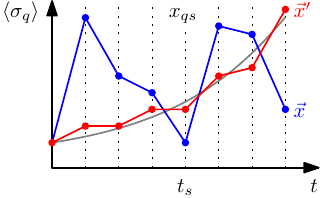}
\caption{Schematic representation of a configuration $\va{x}$ (blue) over all time points. The $\va{x}'$ (red) configuration approximates the real-time dynamics of $\ev{\sigma_q}$ (gray).}
\label{fig:mesh}
\end{figure}

To understand how we employ the results of Sec.~\ref{sec:bbgky} for the purposes of quantum mitigation, we begin by considering the exact dynamics of the spin chain as the solutions of the BBGKY hierarchy of equations.

Start by observing that, for a given $q$, all points $\qty{(t_s, \bar{x}_{qs})}_{s=1, \dots, N_\text{T}}$ lie on a grid of $N_\text{T}(N_\text{S} + 1)$ vertices with, respectively, horizontal and vertical spacings $\Delta t$ and $\Delta x$. This motivates the definition of the $\mathbb{R}^{N_\text{T} \abs{\mathcal{Q}_{r+1}}}$ vector
\begin{equation}
    \va{x} := ((x_{11},\dots,x_{1N_\text{T}}), \dots, (x_{\abs{\mathcal{Q}_{r+1}} 1},\dots,x_{\abs{\mathcal{Q}_{r+1}} N_\text{T}})),
\end{equation}
which we call configuration. The vector $\va{x}$ contains $\abs{\mathcal{Q}_{r+1}}$ sub-vectors (one for each quantity labeled by $q$) of size $N_\text{T}$ (one component for each time point labeled by $s$). While all $\bar{x}_{qs}$ remain fixed, all $x_{qs} \in \mathbb{R}$ are free to vertically vary with respect to the grid. Figure \ref{fig:mesh} depicts the above construction.

If $\va{x}$ were a genuine representation of the true dynamics, then it should approximately satisfy all BBGKY equations of the subhierarchy $\mathcal{Q}_R$, provided they are brought to the discrete grid. To do that, we approximate the time derivative of \eqref{eq:bbkgy_equation} at time $t_s$ ($s=0$ included) as a combination of backward, forward, and centered finite difference schemes
\begin{equation}\label{eq:discrete_derivative}
    \qty[\dv{t} \ev{\sigma_q}]_{t=t_s} \approx \frac{1}{\Delta t}\begin{cases}
        x_{q1} - \bar{x}_{q0}, &\text{if $s = 0$}\\
        (x_{q2} - \bar{x}_{q0})/2, &\text{if $s = 1$}\\
        x_{qN_\text{T}} - x_{q,N_\text{T}-1}, &\text{if $s = N_\text{T}$}\\
        (x_{q,s+1} - x_{q,s-1})/2, &\text{otherwise}
    \end{cases}.
\end{equation}
Specifically, the forward scheme is used at $t_0$, the backward scheme is used at $t_{N_\text{T}}$, and the centered scheme is used at all time-points in-between. Moreover, the approximation at $t_0$ and $t_1$ is special in that it uses the fixed boundary conditions $\bar{x}_{q0}$. After moving all terms to the RHS of \eqref{eq:bbkgy_equation}, we denote the obtained approximation of \eqref{eq:bbkgy_equation} at time $t_s$ as $E_{A}^{a}(s, \va{x})$. Let $\va{\chi}$ be the equilibrium configuration satisfying $E_{A}^{a}(s, \va{\chi}) = 0$ for all $s\in\qty{0,\dots,N_\text{T}}$ and all $(A,a) \in \mathcal{Q}_R$, with the $\chi_{q0} = \bar{x}_{q0}$ initial conditions. Then one can show, as we do in Appendix \ref{app:action_derivation}, that
\begin{equation}\label{eq:exp_val_chi}
    \chi_{qs} = \lim_{\lambda \to \infty} \mathbb{E}_{\va{X} \sim \mathcal{D}(\lambda)}[(\va{X})_{qs}],
\end{equation}
where $\va{X}$ is a random vector, $\mathcal{D}(\lambda)$ is a $N_\text{T} \abs{\mathcal{Q}_{r+1}}$-dimensional distribution with density function (PDF)
\begin{equation}\label{eq:montecarlo_weight}
    \omega(\lambda, \va{x}) := \frac{e^{-\lambda S(\va{x})}}{\int \dd{\va{x}'} e^{-\lambda S(\va{x}')}},
\end{equation}
and $\lambda$ is a parameter which we will later understand as an inverse temperature \cite{Kirkpatrick:1983zz}. The action $S(\va{x})$, which we will generalize in the upcoming Sec.~\ref{sec:our_scheme}, is here only determined by the BBGKY action
\begin{equation}\label{eq:bbgky_action}
    S_\text{B}(\va{x}) := \frac{\abs{\mathcal{Q}_{r+1}}}{\abs{\mathcal{Q}_{r}}} \Delta t \sum_{(A,a) \in \mathcal{Q}_r} \sum_{s=0}^{N_\text{T}} E_A^a(s, \va{x})^2,
\end{equation}
so $S(\va{x}) = S_\text{B}(\va{x})$ where $S_\text{B}(\va{x})$ is evaluated at $r=R$. The BBGKY action is given by the summation of $\abs{\mathcal{Q}_r} (N_\text{T}+1)$ terms: in anticipation of its combination with \eqref{eq:quantum_action} in the generalization \eqref{eq:action} of Sec.~\ref{sec:our_scheme}, \eqref{eq:quantum_action} being the summation of $\abs{\mathcal{Q}_{r+1}} (N_\text{T}+1)$ terms instead, we normalize $S_\text{B}(\va{x})$ with the $\abs{\mathcal{Q}_{r+1}}/\abs{\mathcal{Q}_{r}}$ factor. Effectively, \eqref{eq:exp_val_chi} is the $\lambda \to \infty$ limit of a discretized path integral \cite{Feynman:1948ur, feynman2010quantum}, which we numerically verify in the $r=R$ results of Sec.~\ref{sec:numerical_results}. As a last remark, while \eqref{eq:exp_val_chi} is a universal statement, the distribution $\mathcal{D}(\lambda)$ is a direct consequence of our parametrization of Dirac delta distributions performed in Appendix \ref{app:action_derivation}. Hence, a different choice of such a parametrization would lead to a different $\mathcal{D}(\lambda)$ distribution, without altering the result \eqref{eq:exp_val_chi}.

\subsection{Our mitigation scheme}\label{sec:our_scheme}

Equation \eqref{eq:exp_val_chi} provides an approximation of the exact real-time dynamics as the $\lambda \to \infty$ limit of a classical expectation value, stemming from the distribution $\mathcal{D}(\lambda)$. Since $\mathcal{D}(\lambda)$ is completely determined by a choice of an action in the PDF \eqref{eq:montecarlo_weight}, $S(\va{x})$ has the capability to encode the exact dynamics of the spin chain. As a result, real-time dynamics can be converted into an action form. It is in this spirit that we propose to gather all quantum simulation measurements $\bar{x}_{qs}$ within the quantum action
\begin{equation}\label{eq:quantum_action}
    S_\text{Q}(\va{x}) := \frac{\Delta t}{2} \sum_{q=1}^\abs{\mathcal{Q}_{r+1}} \sum_{s=0}^{N_\text{T}} \qty(\frac{x_{qs} - \bar{x}_{qs}}{y_{qs}})^2,
\end{equation}
where
\begin{equation}\label{eq:variance}
    y_{qs} := 1 - \begin{cases}
    \bar{x}^2_{qs}, &\text{if $\abs{\bar{x}_{qs}} < 1$}\\
    \displaystyle\qty(\frac{N_\text{S} \bar{x}_{qs} - \text{sgn}(\bar{x}_{qs})}{N_\text{S}+1})^2, &\text{if $\abs{\bar{x}_{qs}} = 1$}
    \end{cases}.
\end{equation}
The above is the (modified) standard deviation associated to the set of single-shot measurements composing $\bar{x}_{qs}$, whose definition and use we justify in Appendix \ref{app:z_measurements}. Once exponentiated, the quantum action \eqref{eq:quantum_action} functions as a collection of Gaussian penalty terms, so that if one sets $S(\va{x}) = S_\text{Q}(\va{x})$ then one trivially recovers $\chi_{qs} = \bar{x}_{qs}$. Indeed, \eqref{eq:quantum_action} effectively substitutes all $E^a_A(s, \va{x})$ of \eqref{eq:bbgky_action} with constraints of the form $x_{qs} - \bar{x}_{qs} = 0$.

If all quantum measurements $\bar{x}_{qs}$ were noiseless, then \eqref{eq:quantum_action} would encode the same physical information as the exponentially large sum in \eqref{eq:bbgky_action}, evaluated at $r=R$. However, $\bar{x}_{qs}$ measurements are obtained from noisy quantum devices, undermining the capability of \eqref{eq:quantum_action} to grasp the exact dynamics. To tackle the issue, we propose to generalize the action as
\begin{equation}\label{eq:action}
    S(\va{x}) := \qty(1 - z) S_\text{Q}(\va{x}) + z S_\text{B}(\va{x}),
\end{equation}
where $0 \leq z \leq 1$, which has yet to be defined, interpolates between the two actions, $z=0$ corresponding to $S(\va{x}) = S_\text{Q}(\va{x})$ and $z=1$ to $S(\va{x}) = S_\text{B}(\va{x})$. Importantly, we evaluate $S_\text{B}(\va{x})$ at a radius $r$ much smaller than $R$. With this new action, $\va{x}$ binds to the noisy measurements in $S_\text{Q}(\va{x})$, while at the same time $\va{x}$ tries to satisfy the selected $\mathcal{Q}_r$ BBGKY equations introduced in $S_\text{B}(\va{x})$. This makes the obtained $\chi_{qs}$ more physical than $\bar{x}_{qs}$, hence a more genuine representation of $\ev{\sigma_q (t_s)}$, because $x_{qs}$ is additionally constrained by the considered BBGKY equations. We therefore propose $\chi_{qs}$ as the mitigation of $\bar{x}_{qs}$, where $z \in [0,1]$ quantifies the proportion of included physics informing the mitigation.

Two possible choices for the definition of $z$ are considered in this paper. The natural $z := \abs{\mathcal{Q}_{r}}/\abs{\mathcal{Q}_R}$ definition requires the determination of $\mathcal{Q}_R$, which is an exponentially large subset whose computation undermines its practical use. To avoid this issue, we opt for the alternative definition $z := \abs{\mathcal{Q}_{r}}/\abs{\mathcal{Q}_{r+1}}$ which, as seen in Sec.~\ref{sec:hierarchical_structure}, can be computed with a $\text{poly}(N_\text{Q})$ amount of classical resources. Hence, because of this last suitable property, we employ this second definition of $z$ for all numerical results in the upcoming Sec.~\ref{sec:results}.

Finally, from \eqref{eq:inequalities}, the relationship between these two choices for $z$ is (the inequalities are no longer strict because we allow $r \in \mathbb{N}_0$)
\begin{equation}\label{eq:inequality_alternative_z}
    0 < \frac{\abs{\mathcal{Q}_{r}}}{\abs{\mathcal{Q}_R}} \leq \frac{\abs{\mathcal{Q}_{r}}}{\abs{\mathcal{Q}_{r+1}}} \leq 1.
\end{equation}
From the above, we see that the relative difference between the two ratios shrinks as $r$ increases, until they become equal at $r = R - 1$, and that both ratios become equal to unity at $r=R$.

\subsection{Sampling of the mitigations}\label{sec:metropolis}
The evaluation of the mitigated configuration \eqref{eq:exp_val_chi} implies calculating the $\lambda \to \infty$ limit (task 1) of a $N_\text{T} \abs{\mathcal{Q}_{r+1}}$-dimensional integral (task 2). Task 2 cannot be tackled with quadrature formulas, as it would require the sampling of $\order{N_\text{S}^{N_\text{T} \abs{\mathcal{Q}_{r+1}}}}$ configurations. To overcome this (classical) computational limitation, one can rely on the Metropolis-Hastings (MH) sampling procedure \cite{Metropolis:1953am, Hastings:1970aa}, which we now briefly present.

From an initial configuration $\va{x}_0$, the MH algorithm produces a Markovian chain of configurations $\va{x}_0 \to \va{x}_1 \to \dots \to \va{x}_m \to \dots \to \va{x}_M$ whose coordinates are denoted by $x^m_{qs}$. These configurations are used in the estimation
\begin{equation}\label{eq:mitigation_result}
    \va{\chi}_\text{MH} := \frac{1}{M_\text{S}} \sum_{m=1}^{M_\text{S}} \va{x}_{M_\text{T} + m \frac{M - M_\text{T}}{M_\text{S}}},
\end{equation}
where $M_\text{S}$ is the amount of (equispaced) selected samples after $M_\text{T}$ thermalization sweeps $\va{x}_m \to \va{x}_{m+1}$. In turn, a sweep is composed of $N_\text{T} \abs{\mathcal{Q}_{r+1}}$ proposals $\va{x}_m \to \va{x}'_m$, where $\va{x}'_m$ differs from $\va{x}_m$ by one randomly shifted chosen coordinate. The proposal is accepted with probability $\min(1, \omega(\lambda, \va{x}')/\omega(\lambda, \va{x})) = \min(1, e^{-\lambda \Delta S(\va{x}', \va{x})})$, where $\Delta S(\va{x}', \va{x}) := S(\va{x}') - S(\va{x})$. We optimize the evaluation of $\Delta S(\va{x}', \va{x})$ by only computing the local variation of the action, given by the difference of the terms in $S(\va{x}')$ and $S(\va{x})$ involved in the proposal. 

Thermalization ensures $\va{x}_{M_\text{T}}$ is sufficiently distant from the arbitrary and unphysical $\va{x}_0$, while the amount of selected samples reduces autocorrelation between selected configurations, improving the statistics \cite{landau2021guide}. The configurations of the Markov chain past the thermalization point approximate random draws of $\va{X} \sim \mathcal{D}(\lambda)$. In practice, these configurations will mostly be physical (in the BBGKY sense) oscillations around the quantum measurements.

Task 1 can be simultaneously solved with task 2 if incorporated in the MH sampling algorithm, which then needs to be generalized. We heuristically achieve this by slowly increasing $\lambda$ at each sweep by a small variation $\Delta \lambda$. That way, by identifying $\lambda$ with an inverse temperature, the MH algorithm becomes a simulated annealing procedure \cite{Kirkpatrick:1983zz}, and the Markov chain gets progressively trapped within the peaking $\mathcal{D}(\lambda)$ distribution as $\lambda$ increases, reproducing the limit process. Algorithm \ref{alg:mc} depicts our implementation of this (modified) MH sampling procedure. There, $\Delta x$ is used as a length scale for the random shifts in the proposals, $\mathcal{U}(0,1)$ denotes the uniform distribution across $[0,1]$, and the appearance of a distribution symbol is understood as a random realization of an associated random variable.

{
\SetKwFor{For}{for}{}{}
\SetKwIF{If}{ElseIf}{Else}{if}{}{else if}{else}{}

\begin{algorithm}
\caption{Generation of the Markov chain}\label{alg:mc}
\KwIn{$\quad\va{x}_0$}
\KwOut{$\,\va{x}_1, \dots, \va{x}_M$}
\BlankLine
\BlankLine
$\lambda \gets 0$\\
\BlankLine
\For{$m \in \qty{1,\dots,M}$}{
    \BlankLine
    $\va{x}_m \gets \va{x}_{m-1}$\\
    \BlankLine
    \For{$q \in \qty{1,\dots,\abs{\mathcal{Q}_{r+1}}}$}{
        \For{$s \in \qty{1,\dots,N_\text{T}}$}{
            \BlankLine
            \tcp{Draw random shift}
            $h \gets \mathcal{N}(0, (\Delta x)^2)$\\
            \BlankLine
            \tcp{Compute local variation}
            $\Delta S \gets -S(\va{x}_m)$\\
            $x^m_{qs} \gets x^m_{qs} + h$\\
            $\Delta S \gets \Delta S + S(\va{x}_m)$\\
            \BlankLine
            \tcp{Attempt proposal}
            \If{$\mathcal{U}(0,1) > \min(1, e^{-\lambda \Delta S})$}{
                \BlankLine
                $x^m_{qs} \gets x^m_{qs} - h$\\
            }
        }
    }
    \BlankLine
    \tcp{Decrease temperature}
    $\lambda \gets \lambda + \Delta \lambda$\\
}
\end{algorithm}
}

\section{Chiral dynamics in the Schwinger model}\label{sec:schwinger}
In this Section, we describe the CME showcased in QCD and its implications in current QCD phenomenology research. We then introduce the Schwinger model and its corresponding CME. Finally, we provide its spin-$\frac{1}{2}$ chain Hamiltonian, together with the encoding of the induced electric current.

\subsection{The chiral magnetic effect}\label{sec:cme_qcd}
Consider approximately massless chiral fermions in a strong external magnetic field. Under these conditions, a chiral imbalance between right- and left-handed fermions, quantified by the chemical potential $\mu_5$, generates an electric current (or correspondingly an electric field) along the magnetic field lines: this phenomenon is called the CME \cite{Fukushima:2008xe, Fukushima:2010vw, Kharzeev:2013ffa, Kharzeev:2023zbo, Kharzeev:2024zzm, Warringa:2008kv}.

On a phenomenological level, following \cite{Kharzeev:2024zzm}, the strong magnetic field polarizes the spins of the positively-charged (negatively-charged) particles along (opposite to) its direction, forcing a spin-momentum lock of the fermions dictated by their chirality, which coincides with their helicity in the massless limit \cite{Warringa:2008kv}. That way, assuming approximately parallel spins and momenta, right- (left-) handed fermions are bound to move along (opposite to) the magnetic field lines. At the same time, right-handed positive and left-handed negative (left-handed positive and right-handed negative) fermions collectively generate an electric current along (opposite to) the magnetic field axis. If the density of right-handed fermions matches the density of left-handed fermions, $\mu_5 = 0$, no charge separation is induced. However, in the event of a positive (negative) chiral imbalance $\mu_5 \neq 0$, a positive (negative) macroscopic current is generated along (opposite to) the external strong magnetic field.


Due to its macroscopic nature, the CME can be experimentally observed \cite{Li:2014bha}. As a result, the CME can be used to probe quantum phenomena otherwise inaccessible, such as topological transitions in color-confined QCD matter \cite{Kharzeev:2023zbo}. Indeed, these transitions come with chiral flips, which induce (at least locally) chiral imbalances \cite{Kharzeev:1998kz}. For instance, taking advantage of the strong magnetic fields generated in relativistic heavy-ion collisions, the CME could be observed in the resulting quark-gluon plasma \cite{Kharzeev:2023zbo}. As the QCD topological transitions violate baryon number conservation, their observation via the CME in the rapidly expanding quark-gluon plasma (mimicking the early Universe) could provide a possible explanation to the baryon asymmetry problem \cite{Sakharov:1967dj}.

\subsection{The CME in the Schwinger model}\label{sec:cme}

The (massive $m > 0$) Schwinger model is a $(1+1)$-dimensional theory describing quantum electrodynamics \cite{PhysRev.128.2425}. Despite its simplicity relative to $(3+1)$-dimensional QCD, it shares similar features such as confinement \cite{Coleman:1975pw} and the sign problem \cite{Gattringer:2015nea}, making it a suitable toy model to, respectively, probe QCD and test quantum simulations of field theories. In particular, the Schwinger model also showcases a phenomenon similar to the CME, upon a chiral transformation of the fermionic fields by a time-dependent angle $\theta$ and in the vanishing coupling constant limit \cite{PhysRevResearch.2.023342}. While the former induces the chiral imbalance via the chiral chemical potential $\mu_5 = -\dot{\theta}/2$, the latter decouples the theory into free fermions of chirally-rotated masses \cite{PhysRevResearch.2.023342}. In this particular regime of the Schwinger model, a chiral imbalance induces an electric current as in the CME. Choosing the particular time-dependence for the sudden chiral quench
\begin{equation}
    \dot{\theta}(t) = \begin{cases}
        0, &\text{if $t \leq 0$}\\
        -2\mu_5, &\text{if $t > 0$}
    \end{cases},
\end{equation}
with $\theta(0) = 0$ and $\mu_5$ from now on considered constant, it can be shown that the $t \ll 1$ behaviour of the (spatially averaged) induced electric current is \cite{PhysRevResearch.2.023342}
\begin{equation}\label{eq:cme}
    \ev{j}_\text{avg} = 4\mu_5 m t^2 \ev{\bar{\psi}\psi}_\text{avg} + \order{t^3},
\end{equation}
where $\ev{\bar{\psi}\psi}_\text{avg} \sim -m$ is the (spatially averaged) chiral condensate.

The discretized lattice formulation of the Schwinger model, in the above particular form, is obtained using Kogut-Susskind fermions \cite{Kogut:1974ag}. Then, with an application of the Jordan-Wigner transformation \cite{Jordan:1928wi}, the lattice is mapped to a quantum spin-$\frac{1}{2}$ chain. By forcing periodic boundary conditions, the $N_\text{Q} \in 2 \mathbb{N}_1$ qubits Hamiltonian of the model brought to the spin chain reads \cite{PhysRevResearch.2.023342}
\begin{equation}\label{eq:cme_H}
\begin{split}
    H &=\sum_{k=0}^{\frac{N_\text{Q}}{2} - 1} \Bigl[ a_k \qty(\sigma^1_{2k+1} \sigma^1_{2k+2} + \sigma^2_{2k+1} \sigma^2_{2k+2})\\
    &- \frac{\dot{\theta}}{8} \qty(\sigma^1_{2k+1} \sigma^2_{2k+2} - \sigma^2_{2k+1} \sigma^1_{2k+2}) \Bigl]\\
    &+ \sum_{k=1}^{\frac{N_\text{Q}}{2} - 1} \Bigl[ a_k \qty(\sigma^1_{2k} \sigma^1_{2k+1} + \sigma^2_{2k} \sigma^2_{2k+1})\\
    &- \frac{\dot{\theta}}{8} \qty(\sigma^1_{2k} \sigma^2_{2k+1} - \sigma^2_{2k} \sigma^1_{2k+1}) \Bigl]\\
    &+ (-1)^\frac{N_\text{Q}}{2} a_{N_\text{Q} - 1} \qty(\sigma^1_1 \sigma^1_{N_\text{Q}} + \sigma^2_1 \sigma^2_{N_\text{Q}}) \prod_{k=2}^{N_\text{Q} - 1} \sigma^3_k\\
    &- (-1)^\frac{N_\text{Q}}{2} \frac{\dot{\theta}}{8} \qty(\sigma^2_1 \sigma^1_{N_\text{Q}} - \sigma^1_1 \sigma^2_{N_\text{Q}}) \prod_{k=2}^{N_\text{Q} - 1} \sigma^3_k\\
    &- \frac{m}{2}\cos(\theta) \sum_{k=1}^{N_\text{Q}} (-1)^k \sigma^3_k\\
    &=: \sum_{(B,b) \in \mathcal{H}} h^b_B(t) \sigma^b_B,
\end{split}
\end{equation}
where $a_k := \frac{1}{2}\qty[\omega - (-1)^k \frac{m}{2} \sin(\theta)]$ and $\omega$ is half of the inverse lattice spacing. Notice that $L_\mathcal{H} = N_\text{Q}$, but $\abs{\mathcal{H}} \sim \text{poly}(N_\text{Q})$. Likewise, the quantum observable corresponding to the (spatially averaged) electric current \eqref{eq:cme} brought to the spin chain reads \cite{PhysRevResearch.2.023342}
\begin{equation}\label{eq:electric_current}
\begin{split}
    J &:= \frac{\omega}{2 N_\text{Q}} \sum_{k=0}^{\frac{N_\text{Q}}{2} - 1} \qty(\sigma^1_{2k+1} \sigma^2_{2k+2} - \sigma^2_{2k+1} \sigma^1_{2k+2})\\
    &+ \frac{\omega}{2 N_\text{Q}} \sum_{k=1}^{\frac{N_\text{Q}}{2} - 1} \qty(\sigma^1_{2k} \sigma^2_{2k+1} - \sigma^2_{2k} \sigma^1_{2k+1})\\
    &+ (-1)^\frac{N_\text{Q}}{2} \frac{\omega}{2 N_\text{Q}} \qty(\sigma^2_1 \sigma^1_{N_\text{Q}} - \sigma^1_1 \sigma^2_{N_\text{Q}}) \prod_{k=2}^{N_\text{Q} - 1} \sigma^3_k\\
    &=: \sum_{q=1}^{\abs{\mathcal{Q}_0}} J_q \sigma_q,
\end{split}
\end{equation}
where we take $\mathcal{Q}_0$ to be the set of all Pauli strings $\sigma_q$ appearing in $J$, and where by $J_q$ we denote the real coefficient associated to $\sigma_q$. Again, we verify that $\abs{\mathcal{Q}_0} \sim \text{poly}(N_\text{Q})$. The above is the main object of our study, namely \eqref{eq:electric_current} is the quantity of interest whose quantum simulations we wish to mitigate.

\section{Results of the mitigation}\label{sec:results}

In this Section, we empirically assess the effectiveness of our proposed mitigation scheme at recovering the CME dynamics from noisy simulations of the Schwinger model. We explain the setting under which quantum simulations are performed and we introduce the evaluation metrics. We then provide the numerical results and discuss them. Finally, we study the structure of the corresponding BBGKY hierarchy.

\subsection{Numerical framework}\label{sec:simulations}
We use Qiskit $2.1$ \cite{qiskit2024} to perform Trotterized simulations with Hamiltonian \eqref{eq:cme_H} on a classical emulation of a quantum device, affected by attenuated realistic quantum error. The latter is implemented through a noise model compiled from the physical backend properties of the IBM Torino QPU \footnote{All simulations were performed using the online noise model snapshot of September 18th 2025.}. As the current intensity of NISQ error is too great for $N_\text{Q} \geq 6$ simulations, we attenuate the effect of the noise model by 90\% in the way explained in Appendix \ref{app:intensity_noise}, thereby affecting quantum simulations only by 10\% of the realistic quantum noise currently generated in IBM's QPUs.

Simulations begin with an initial state corresponding to the ground state (or, in case of degeneracy, to a uniform superposition of all ground states) of the Hamiltonian still unaffected by the quench, found by exact diagonalization of $H(0)$. Then, for the remainder of all quantum simulations, we set $\omega = 1$, $T=3$, $N_\text{Q} = 8$, $N_\text{S} = 10^4$ and $N_\text{T} = 10$. For the execution of the mitigation procedure, we set $M = 10^4$, $M_\text{T} = M/4 = 2500$, $M_\text{S} = 30$ (hence the sampling separation is of $(M - M_\text{T})/M_\text{S} = 250$ configurations) and, based on numerical experiments, we found $\Delta \lambda = 1$ to be sufficiently small to successfully perform the simulated annealing procedure from $\lambda = 0$ to the large $\lambda = M \Delta \lambda$. The Markov chain begins with $\va{x}_0$ as a random hot start (all $x^0_{qs}$ are randomly distributed across $[-2,2]$), leading to numerical results which are consistent with an ergodic sampling. The unspecified parameters $m$, $\mu_5$ and $r$ are varied across simulations.

Estimations of the electric current \eqref{eq:electric_current} are numerically computed as
\begin{equation}\label{eq:various_measurements}
\begin{split}
    \ev{J(t_s)}_\text{Noisy} &:= \sum_{q=1}^\abs{\mathcal{Q}_0} J_q \bar{x}_{qs},\\
    \ev{J(t_s)}_\text{MH} &:= \sum_{q=1}^\abs{\mathcal{Q}_0} J_q \chi_{qs}.
\end{split}
\end{equation}

Moreover, with the purpose of comparing the \eqref{eq:various_measurements} results with the ideal dynamics and, as we explain below, to assess their Trotter error, we introduce classical exact diagonalization (ED) estimations of the electric current, obtained with the \eqref{eq:time_evolution_operator} approximation, computed as
\begin{equation}
\begin{split}
    \ev{J(t_s)}_\text{ed} &:= \sum_{q=1}^\abs{\mathcal{Q}_0} J_q \ev{\sigma_q(t_s)}_\text{ed},\\
    \ev{J(t_s)}_\text{ED} &:= \sum_{q=1}^\abs{\mathcal{Q}_0} J_q \ev{\sigma_q(t_s)}_\text{ED},
\end{split}
\end{equation}
where $\ev{J(t_s)}_\text{ed}$ employs $N_\text{T} = 10$ whereas $\ev{J(t_s)}_\text{ED}$ employs $N_\text{T} = 100$ (however $s$ keeps referring to the $N_\text{T} = 10$ discretization of time). Indeed, the type of errors affecting all of the estimates are:
\begin{itemize}
    \item[] $\ev{J(t_s)}_\text{ed/ED}$ (composed of $\ev{\sigma_q(t_s)}_\text{ed/ED}$)
    \begin{itemize}
        \item ED computation of the \eqref{eq:time_evolution_operator} approximation
        \item Affected by: Trotter error
    \end{itemize}
    
    \item[] $\ev{J(t_s)}_\text{Noisy}$ (composed of $\bar{x}_{qs}$)
    \begin{itemize}
        \item Realistic execution of the noisy quantum circuit implementing \eqref{eq:time_evolution_operator}
        \item Affected by: Trotter error, shot noise and quantum noise
    \end{itemize}
    
    \item[] $\ev{J(t_s)}_\text{MH}$ (composed of $\chi_{qs}$)
    \begin{itemize}
        \item Result of our mitigation scheme \eqref{eq:mitigation_result}
        \item Affected by: hyperparameter choice in the sampling method
    \end{itemize}
\end{itemize}
To assess the Trotter error of both $\ev{J(t_s)}_\text{ed}$ and $\ev{J(t_s)}_\text{ED}$, we define
\begin{equation}
    L_\text{Trotter} := \sqrt{\Delta t\sum_{s=0}^{N_\text{T}} \biggl(\ev{J(t_s)}_\text{ed} - \ev{J(t_s)}_\text{ED}\biggr)^2}.
\end{equation}
Due to the linear in $\Delta t$ behaviour of the \eqref{eq:time_evolution_operator} Trotter error, $\ev{J(t_s)}_\text{ed}$ and $\ev{J(t_s)}_\text{ED}$ have a Trotter error relative to the exact dynamics $\ev{J(t)}$ of, respectively, approximately $C \Delta t$ and $C \Delta t/10$, where $C > 0$ is an unknown common constant. Then, their absolute difference
$\abs{\ev{J(t_s)}_\text{ed} - \ev{J(t_s)}_\text{ED}}$ is of the order of $(9/10)C\Delta t$. As a result, $2L_\text{Trotter}$ gives an upper bound of the accumulated Trotter error between $\ev{J(t_s)}_\text{ed}$ and $\ev{J(t)}$, and likewise $L_\text{Trotter}$ gives an upper bound of the accumulated Trotter error between $\ev{J(t_s)}_\text{ED}$ and $\ev{J(t)}$. Overall, $L_\text{Trotter}$ allows to quantify the quality of the Trotterization \eqref{eq:time_evolution_operator} compared to the exact dynamics, namely whether $N_\text{T}$ is taken sufficiently large (equivalently $\Delta t$ taken sufficiently small) with respect to the intrinsic exact dynamics timescale.

After making sure that $L_\text{Trotter}$ is small enough, the overall accumulated error of $\ev{J(t_s)}_\text{Noisy}$ and $\ev{J(t_s)}_\text{MH}$ at the $r$ radius of $\mathcal{Q}_r$ is quantified by the norm
\begin{equation}\label{eq:2-norm}
    L_\text{Noisy/MH}^r := \sqrt{\Delta t\sum_{s=0}^{N_\text{T}} \biggl(\ev{J(t_s)}_\text{Noisy/MH} - \ev{J(t_s)}_\text{ED}\biggr)^2}.
\end{equation}
To evaluate the capability of our method to recover the short-time $t_s \leq 1.2$ behaviour of the electric current \eqref{eq:cme}, we also introduce the metric
\begin{equation}\label{eq:p_metric}
    P^r_\text{Noisy/MH} := \sqrt{\frac{\qty(\va{p}_\text{Noisy/MH} - \va{p}_\text{ED})^2}{\va{p}_\text{ED}^2}}.
\end{equation}
In the above, $\va{p}_\text{ED/Noisy/MH} \in \mathbb{R}^2$ are the two coefficients of the degree-2 polynomials (with a null $t^0$ coefficient) fitting their respective $\ev{J(t_s)}_\text{ED/Noisy/MH}$ simulations, performed at radius $r$. The choice of the fitting polynomial stems from the quadratic in time \eqref{eq:cme} short-time behaviour of the CME electric current. Overall, the metric $P^r_\text{Noisy/MH}$ quantifies the relative distance of the computed $\va{p}_\text{Noisy/MH}$ fits from the expected $\va{p}_\text{ED}$ fit, the latter reproducing \eqref{eq:cme} up to Trotter error.

To conclude, note that $P^r_\text{Noisy/MH}$ restricts to short times $t_s \leq 1.2$, whereas $L_\text{Noisy/MH}^r$ considers the totality of the time evolution $t_s \in [0, T]$. Moreover, we understand that $L_\text{Noisy}^r$ and $P^r_\text{Noisy}$, unlike $L_\text{MH}^r$ and $P^r_\text{MH}$, do not depend on $r$ (this will be verified in the upcoming Sec.~\ref{sec:numerical_results})

\subsection{Numerical results}\label{sec:numerical_results}

In the following, all error bars are computed with the standard propagation of errors formula, truncated at leading order. The propagation of errors for \eqref{eq:various_measurements}, \eqref{eq:2-norm} and \eqref{eq:p_metric} is performed from the individual errors associated to $\bar{x}_{qs}$ and $\chi_{qs}$, which are respectively the standard deviation of the $N_\text{S}$ measurements averaging to $\bar{x}_{qs}$, and the standard deviation of the $M_\text{S}$ selected configurations averaging to $\chi_{qs}$.

\begin{figure}
\centering
\includegraphics[width=\linewidth]{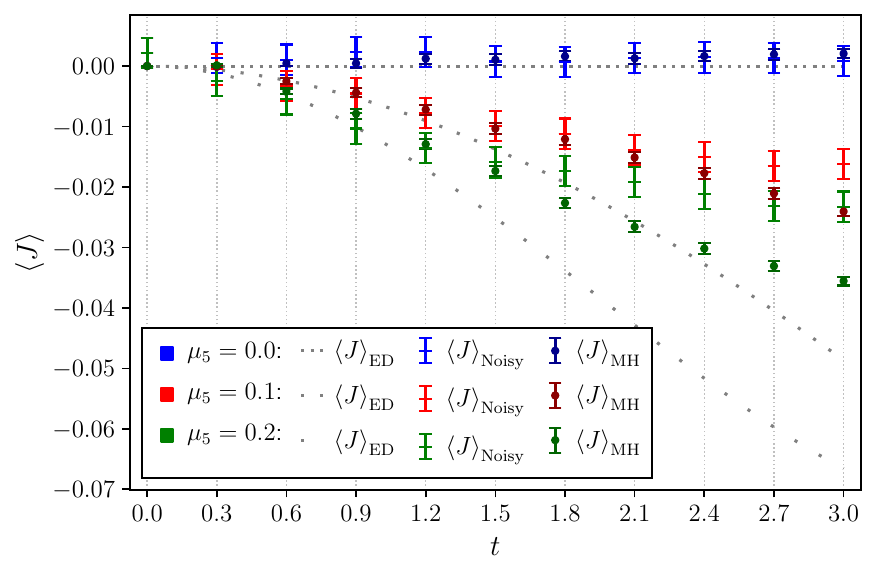}
\caption{Mitigation results of the electric current evolution in time, with $r = 1$ ($R=3$) and $m = 0.1$. Each color corresponds to a different choice of $\mu_5$.}
\label{fig:resultsa}
\end{figure}
\begin{figure}
\centering
\includegraphics[width=\linewidth]{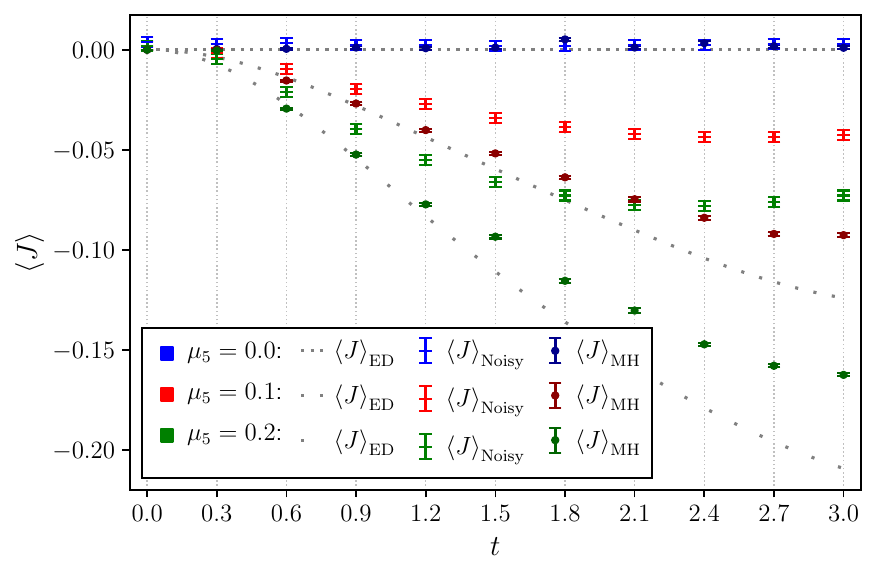}
\caption{Mitigation results of the electric current evolution in time, with $r = 2$ ($R=3$) and $m = 0.5$. Again each color corresponds to a different choice of $\mu_5$.}
\label{fig:resultsb}
\end{figure}

Figures \ref{fig:resultsa} and \ref{fig:resultsb} show the measured estimations of the electric current \eqref{eq:various_measurements} across time, including ED simulations, noisy runs of the quantum simulations, and the mitigations obtained with our method. These are performed at different $m$ and $\mu_5$ values, or Schwinger model realizations, and with different radii $r$. Namely, both Fig.~\ref{fig:resultsa} and \ref{fig:resultsb} simulations use the same set of different chemical potentials, but they differ in the chosen mass and $\mathcal{Q}_r$ radius. We see that, in both Figures, the ED simulations correctly depict the quadratic \eqref{eq:cme} short-time behaviour of the current, while no such feature of the CME can be clearly observed in both noisy measurements. This later fact is corroborated by the $P^r_\text{Noisy}$ results of Table \ref{tbl:table}, which includes the values of the $P^r_\text{Noisy/MH}$ metric for the $\mu_5 \neq 0$ simulations of Fig.~\ref{fig:resultsa} and \ref{fig:resultsb}. In both Figures we see that our mitigation scheme manifestly reduces the errors of the CME simulations, with a clear improvement in Fig.~\ref{fig:resultsb}. In particular, we see in Fig.~\ref{fig:resultsb} that our scheme is also manifestly able to recover the quadratic in time short-time behaviour of the current, which will again be confirmed by the results of Table \ref{tbl:table}.

\begin{figure}
\centering
\includegraphics[width=0.9\linewidth]{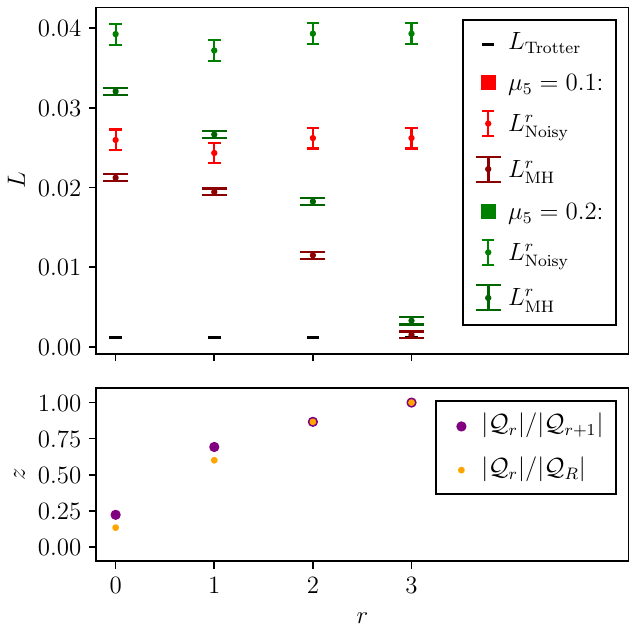}
\caption{Top panel: dependence of the errors with respect to the radius $r$ in the $m = 0.1$ realization. The two colors represent the two different choices of $\mu_5 \neq 0$, for which the corresponding two $L_\text{Trotter}$ overlap. Bottom panel: dependence of the two definitions of $z$ in $r$. The subhierarchy radius is $R=3$.}
\label{fig:a_1_2}
\end{figure}
\begin{figure}
\centering
\includegraphics[width=0.9\linewidth]{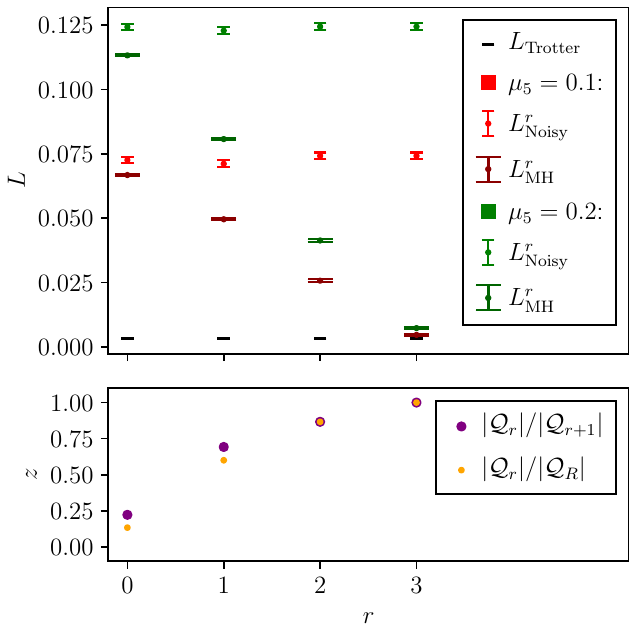}
\caption{Top panel: dependence of the errors with respect to the radius $r$ in the $m = 0.5$ realization. Again, the two colors represent the two different choices of $\mu_5 \neq 0$, for which the corresponding two $L_\text{Trotter}$ yet again overlap. Bottom panel: dependence of the two definitions of $z$ in $r$. The subhierarchy radius is $R=3$.}
\label{fig:b_1_2}
\end{figure}

Figures \ref{fig:a_1_2} and \ref{fig:b_1_2} precisely quantify the errors of Fig.~\ref{fig:resultsa} and \ref{fig:resultsb} over the ED evolution at different radii $r$, while also showing the different $z$ values employed in the mitigations. Specifically, Fig.~\ref{fig:a_1_2} refers to the simulations displayed in Fig.~\ref{fig:resultsa}, whereas Fig.~\ref{fig:b_1_2} refers to the simulations of Fig.~\ref{fig:resultsb}. First of all, in all three Figures, the small values of the accumulated Trotter error $L_\text{Trotter}$ allow us to confirm that our ED simulations are sufficiently close to the exact dynamics \cite{PhysRevResearch.2.023342}, so that the use of the 2-norm \eqref{eq:2-norm} is valid. Moreover, we see that all $L^r_\text{Noisy}$ are independent of $r$: this is expected because the noisy results are not mitigated, hence the choice of $\mathcal{Q}_r$ does not have an effect on $L^r_\text{Noisy}$. Then, we see in Fig.~\ref{fig:a_1_2} and \ref{fig:b_1_2} that not only all $L^r_\text{MH}$ are systematically lower than all $L^r_\text{Noisy}$, rendering our mitigation scheme effective for any choice of radius $r$, but we also see that all $L^r_\text{MH}$ decrease as $r$ approaches $R$, which in these two Figures is $R=3$. Indeed, we previously predicted with \eqref{eq:inequality_alternative_z} that the relative difference between the two choices of $z$ is null at $r=R-1$, which we observe here at $r=2$. The improvement of $L^r_\text{MH}$ as $r$ approaches $R$ is expected because, the more physical information is included is the action \eqref{eq:action}, the more physically constrained the mitigations will be. In particular, in the extreme $r=R$ case corresponding to $z=1$, $S_\text{Q}(\va{x})$ is completely suppressed in favor of $S_\text{B}(\va{x})$, making our mitigation method a purely classical numerical scheme. This last point explains why $L^R_\text{MH}$ is of the same order as $L_\text{Trotter}$. In both Figures, we notice that both definitions of the $z$ parameter grow similarly with radius $r$, justifying the use of $z = \abs{\mathcal{Q}_{r}}/\abs{\mathcal{Q}_{r+1}}$ over $z = \abs{\mathcal{Q}_{r}}/\abs{\mathcal{Q}_{R}}$ in the mitigations. Moreover, the former definition of $z$ is justified a posteriori by the observed decreasing of $L^r_\text{MH}$ as $r$ increases towards $R$. Finally, we verify that \eqref{eq:inequality_alternative_z} is always satisfied, and that $\abs{\mathcal{Q}_{r}}/\abs{\mathcal{Q}_{r+1}}$ approaches $\abs{\mathcal{Q}_{r}}/\abs{\mathcal{Q}_{R}}$ as $r$ increases. An initial growth of $z$ around $r=0$ is followed by a flattening towards $r=R$, in accordance with the subhierarchy propagation depicted in Fig.~\ref{fig:hierarchy}, which we discuss below.

\renewcommand{\arraystretch}{1.2}
\begin{table}
\begin{ruledtabular}
\begin{tabular}{lll|ll}
$r$ & $m$ &
$\mu_5$ &
$P^r_\text{Noisy}$ &
$P^r_\text{MH}$ (Ours)\\[2pt]
\colrule
$0$ & $0.1$ & $0.1$ & $0.532 \pm 1.015$ & \textbf{0.275 $\pm$ 0.074}\\
$1$ & $0.1$ & $0.1$ & $0.557 \pm 1.021$ & \textbf{0.318 $\pm$ 0.059}\\
$2$ & $0.1$ & $0.1$ & $0.774 \pm 1.040$ & \textbf{0.255 $\pm$ 0.028}\\
$3$ & $0.1$ & $0.1$ & $0.774 \pm 1.040$ & \textbf{0.093 $\pm$ 0.009}\\
\colrule
$0$ & $0.1$ & $0.2$ & $0.844 \pm 0.554$ & \textbf{0.310 $\pm$ 0.028}\\
$1$ & $0.1$ & $0.2$ & $0.942 \pm 0.557$ & \textbf{0.235 $\pm$ 0.015}\\
$2$ & $0.1$ & $0.2$ & $0.233 \pm 0.562$ & \textbf{0.148 $\pm$ 0.017}\\
$3$ & $0.1$ & $0.2$ & $0.233 \pm 0.562$ & \textbf{0.075 $\pm$ 0.020}\\
\colrule
$0$ & $0.5$ & $0.1$ & $0.851 \pm 0.236$ & \textbf{0.328 $\pm$ 0.004}\\
$1$ & $0.5$ & $0.1$ & $0.916 \pm 0.236$ & \textbf{0.287 $\pm$ 0.007}\\
$2$ & $0.5$ & $0.1$ & $0.477 \pm 0.234$ & \textbf{0.165 $\pm$ 0.007}\\
$3$ & $0.5$ & $0.1$ & $0.477 \pm 0.234$ & \textbf{0.143 $\pm$ 0.006}\\
\colrule
$0$ & $0.5$ & $0.2$ & $0.797 \pm 0.126$ & \textbf{0.375 $\pm$ 0.005}\\
$1$ & $0.5$ & $0.2$ & $0.821 \pm 0.126$ & \textbf{0.242 $\pm$ 0.009}\\
$2$ & $0.5$ & $0.2$ & $0.510 \pm 0.125$ & \textbf{0.121 $\pm$ 0.003}\\
$3$ & $0.5$ & $0.2$ & $0.510 \pm 0.125$ & \textbf{0.017 $\pm$ 0.004}\\
\colrule
\end{tabular}
\end{ruledtabular}
\caption{\label{tbl:table}%
Ability of the mitigation scheme at recovering the short-time $t_s \leq 1.2$ behaviour of the CME induced electric current, using the $P^r_\text{Noisy/MH}$ metric between fits of ED, noisy and MH results. The row groupings contain simulations of a common Schwinger model realization, mitigated at different radii $r$. The 2nd and 6th row correspond to the simulations of Fig.~\ref{fig:resultsa}, whereas the 11th and the 15th row correspond to the simulations of Fig.~\ref{fig:resultsb}. The subhierarchy radius is $R = 3$.
}
\end{table}

Table \ref{tbl:table} quantifies the capacity of our method at recovering the quadratic in time behaviour of the electric current at short times $t_s \leq 1.2$, by comparing the $P^r_\text{Noisy}$ metric against $P^r_\text{MH}$ on different Schwinger model realizations. We verify that $r$ does not have an effect on $P^r_\text{Noisy}$. The apparent variation of the $P^r_\text{Noisy}$ results is only due to a different random number generator employed across different simulations. Then, on all rows of Table \ref{tbl:table}, we see that $P^r_\text{MH}$ is systematically lower than $P^r_\text{Noisy}$, confirming the previous observation that our mitigation scheme is effective at any radius $r$. Indeed, if all $L^r_\text{MH}$ are smaller than $L^r_\text{Noisy}$, then it is expected that all $P^r_\text{MH}$ will be smaller than $P^r_\text{Noisy}$, since $P^r_\text{Noisy/MH}$ fits the same measurements employed in the computation of $L^r_\text{Noisy/MH}$. Furthermore, we notice that the errors associated to $P^r_\text{MH}$ are lower than those associated to $P^r_\text{Noisy}$. The noisy polynomial fits are therefore highly uncertain, whereas the mitigated fits are more robust, reflecting the recovery of a proper quadratic in time short-time electric current behaviour. Finally we see that, in all considered Schwinger model realizations, $P^{r+1}_\text{MH}$ is always smaller (up to errors) than $P^r_\text{MH}$. This is because, equivalently to the Fig.~\ref{fig:a_1_2} and \ref{fig:b_1_2} discussions, a bigger portion of the subhierarchy constraints the configurations sampling as $r$ increases towards $R$, hence more physical information affects the mitigation.

\subsection{Visualization of the subhierarchy}\label{sec:hierarchy_visualization}

\begin{figure}
\centering
\includegraphics[width=\linewidth]{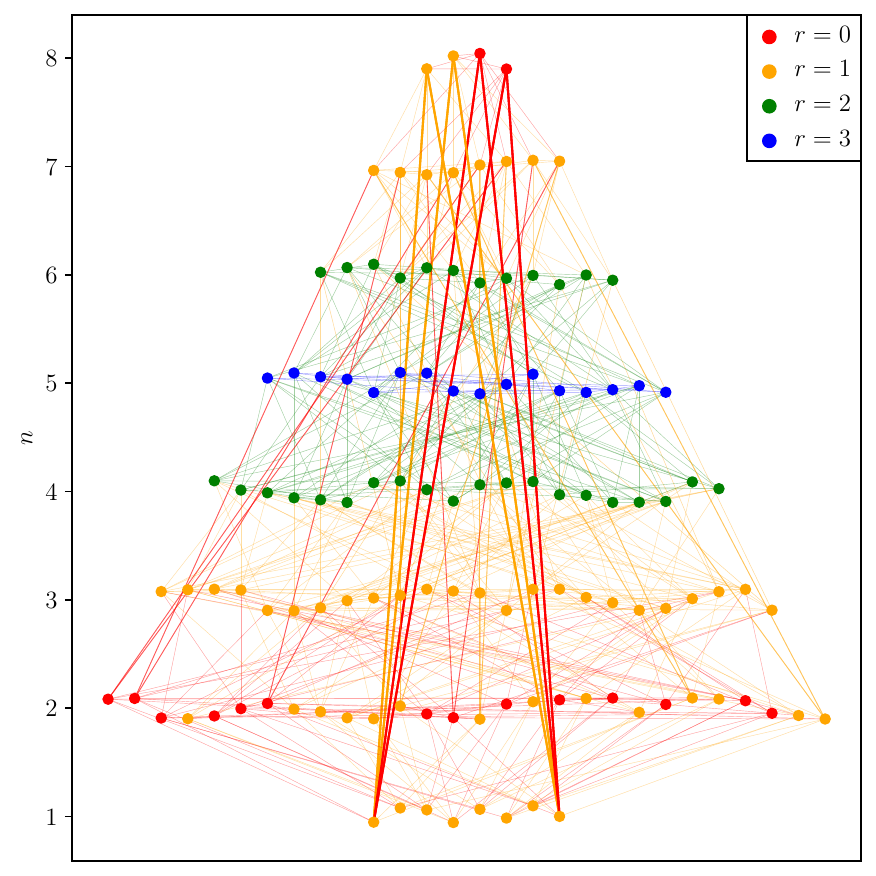}
\caption{Expansion $\mathcal{Q}_{r\in\qty{0,1,2,3}}$ of the subhierarchy $\mathcal{Q}_3$ associated to the Pauli strings of the electric current $\mathcal{Q}_0$ when $\mu_5 \neq 0$. Each node of the graph represents a correlator, where $n$ is its length, and each edge of the graph represents an immediate connection. The radius $r$ is the number of immediate connections separating any correlator to one quantity of $\mathcal{Q}_0$. The color of an edge corresponds to the color of the smallest $r$ between the two immediately connected nodes. The width of an edge is proportional to the absolute difference between the lengths of the two immediately connected nodes.}
\label{fig:hierarchy}
\end{figure}
Finally, Fig.~\ref{fig:hierarchy} shows the $r \in \qty{0,\dots,R}$ growth of the subhierarchy associated to the Pauli strings $\mathcal{Q}_0$, when $\mu_5 \neq 0$. We verify that $\mathcal{Q}_3 = \mathcal{Q}_4$, hence the subhierarchy radius is $R=3$, and we also verify that all elements of $\mathcal{Q}_0$ are connected with each other. Numerically we found that $\mathcal{Q}_R$ contains $120$ BBGKY equations and is one among the $17$ independent subhierarchies splitting the full hierarchy of $4^{N_\text{Q}} = 65536$ equations. By looking at the immediate connections, Fig.~\ref{fig:hierarchy} also allows us to verify the validity of the \eqref{eq:length_bounds} bounds. Indeed, the $\abs{B}\in\qty{1,2}$ terms in $\mathcal{H}$ force $n$-point correlators to only immediately connect to $(n-1)$-point, $n$-point and $(n+1)$-point correlators. This is observed in the expansion of the nodes coloring, which proceeds by neighbour layers of correlator length. The four $\abs{B} = N_\text{Q}$ interaction terms in $\mathcal{H}$ force $n$ point correlators to only immediately connect to correlators whose length is greater than or equal to $N_\text{Q} - (n-1)$. This can be observed for instance in the particular case of the immediate connections of $\mathcal{Q}_0$. Indeed, with our Hamiltonian \eqref{eq:cme_H}, while the 2-point correlators of $\mathcal{Q}_0$ do immediately connect to 7-point correlators, they do not immediately connect to 8-point correlators, even if allowed by \eqref{eq:length_bounds}. Moreover, these 8-point correlators of $\mathcal{Q}_0$ are permitted to immediately connect to any other correlator, but only do so to 1-point, 7-point and 8-point correlators, as it is clearly visible by, respectively, the vertical thick, vertical thin and horizontal thin red immediate connections of the 8-point correlators in Fig.~\ref{fig:hierarchy}. Finally, we notice that the hierarchical growth of $\mathcal{Q}_r$ is reflected in the increase of $z$ in all bottom panels of Fig.~\ref{fig:a_1_2} and \ref{fig:b_1_2}, that is, an initial growth around $z=0$ followed by a flattening towards $z=1$.

\section{Conclusions}\label{eq:conclusions}
In this paper, we introduced an extended qubit BBGKY hierarchy, employed it to formulate a novel quantum error mitigations scheme, and applied the latter on noisy quantum simulations of the Schwinger model CME. More precisely, we computed qubit BBGKY equations which are able to accommodate for arbitrary-many qubits time-dependent Hamiltonians. This in turn allowed us to get a deeper understanding of the corresponding BBGKY hierarchy structure.

We showed that, provided the Hamiltonian is built with a $\text{poly}(N_\text{Q})$ amount of Pauli strings and that the initial state is a linear combination of a polynomially-large amount of computational quantum basis states, that the classical and quantum resources required to evaluate the BBGKY equations and to determine their immediate connections throughout the hierarchy is polynomial in $N_\text{Q}$, motivating the applicability of the BBGKY hierarchy for quantum error mitigation purposes. This led to the formulation of our method, where configurations representing possible mitigations are sampled by a simulated annealing algorithm, under the guidance of noisy quantum measurements and of physical constraints provided by a selected portion of the BBGKY hierarchy.

To assess the effectiveness of our scheme, we compared the physics-informed mitigations with the exact dynamics of the Schwinger model CME. We empirically demonstrated a systematic improvement of all performed CME quantum simulations, regardless of the radius $r$, both in terms of global noise reduction and of recovery of short-time dynamical behaviour. Moreover, we numerically showed that the effectiveness of our method at reducing quantum noise in both cases scales with $r$. Importantly, this implies that our method is readily applicable even when only a small polynomial portion of the exponentially large subhierarchy is implemented. In the particular case of the Schwinger model Hamiltonian, the theoretically predicted behaviour of the immediate connections among BBGKY equations was verified.

Possible extensions of this work include the application of our method for variational imaginary-time evolution as a post-processing procedure. Finally, our scheme could also mitigate the execution of arbitrary quantum circuits, if translated into Hamiltonian formulations.

\begin{acknowledgments}
We would like to thank Yuta Kikuchi for providing the source code supporting the findings in \cite{PhysRevResearch.2.023342}. This work has received support from the French State managed by the National Research Agency under the France 2030 program with reference ANR-22-PNCQ-0002. We acknowledge the use of IBM Quantum services for this work. The views expressed are those of the authors, and do not reflect the official policy or position of IBM or the IBM Quantum team.
\end{acknowledgments}

\section*{Data Availability}
The data that support the findings of this article are openly available \cite{dataset2}. The source codes that generated and processed the data are available at \href{https://github.com/theosaporiti/BBGKY-ISM}{\texttt{https://github.com/theosaporiti/BBGKY-ISM}}.

\appendix

\section{Glossary of symbols}\label{app:symbols}
\begin{longtable}{lll}
\textbf{Symbol} & \textbf{Meaning} \\
\colrule
\colrule
& \textit{Pauli strings and connections}\\
\colrule
$N_\text{Q}$ & Number of qubits\\
$S$ & System set\\
$A,B,C$ & Subsystem set\\
$i,j,k$ & Site\\
$a,b,c$ & Directions\\
$a_i,b_i,c_i$ & Direction\\
$\sigma_i^{a_i}$ & Pauli operator\\
$\sigma_A^a, \sigma_B^b, \sigma_C^c, \sigma_q$ & Pauli string\\
$(A,a), (B,b), (C,c)$ & Pauli strings' encodings\\
$\sigma^a_A \leftrightarrow \sigma^c_C$ & Immediate connection\\
$\sigma^a_A \leftrightarrow \dots \leftrightarrow \sigma^c_C$ & Connection\\
\colrule
& \textit{Hamiltonian}\\
\colrule
$H$ & Hamiltonian\\
$\mathcal{H}$ & Interaction terms\\
$h_B^b$ & Coupling constant of $\sigma_B^b$\\
$L_\mathcal{H}$ & Maximal Pauli string length in $H$\\
\colrule
& \textit{BBGKY equations}\\
\colrule
$L_{AB}^{ab}, n$ & Length of $\comm{\sigma_A^a}{\sigma_B^b}$\\
$f^{ab}_{AB}$ & Factor of $\comm{\sigma_A^a}{\sigma_B^b}$\\
$d_{AB}^{ab}$ & Number of different directions\\
\colrule
& \textit{BBGKY hierarchy}\\
\colrule
$\mathcal{Q}_0$ & Pauli strings to mitigate\\
$\mathcal{Q}_r$ & Subhierarchy portion\\
$\mathcal{Q}_R$ & Subhierarchy\\
$r$ & Radius\\
$R$ & Subhierarchy radius\\
$z$ & Subhierarchy portion ratio\\
\colrule
& \textit{Trotterization}\\
\colrule
$N_\text{T}$ & Number of Trotter slices\\
$T$ & Total time\\
$t, t_s$ & Time\\
$s$ & Time point\\
$\Delta t$ & Time step\\
$\Delta x$ & Observable measurement spacing\\
\colrule
& \textit{Measurements and configurations}\\
\colrule
$N_\text{S}$ & Number of shots\\
$\bar{x}_{qs}$ & Quantum measurements\\
$\bar{x}_{q0}$ & Boundary conditions\\
$\Tilde{x}_{qs}$ & Noiseless quantum simulations\\
$\chi_{qs}$ & Mitigation results\\
$\va{x}$ & Configuration\\
$\va{x}_m$ & $m$-th sampled configuration\\
$\va{x}'_m$ & Proposal for $\va{x}_m \to \va{x}'_m$\\
$x_{qs}$ & Coordinates of $\va{x}$\\
$x^m_{qs}$ & Coordinates of $\va{x}_m$\\
$y_{qs}$ & Modified standard deviation of $x_{qs}$\\
\colrule
& \textit{BBGKY action}\\
\colrule
$S(\va{x})$ & Action\\
$S_\text{Q}(\va{x})$ & Quantum action\\
$S_\text{B}(\va{x})$ & BBGKY action\\
$E_A^a(s,\va{x})$ & Discretized BBGKY equation of $\sigma_A^a$\\
$\mathcal{D}(\lambda)$ & Multidimensional distribution\\
$\omega(\lambda, \va{x})$ & Probability density function\\
\colrule
& \textit{MH hyperparameters}\\
\colrule
$M$ & Number of sweeps\\
$M_\text{S}$ & Number of samples\\
$M_\text{T}$ & Thermalization steps\\
$\lambda$ & Inverse temperature\\
\colrule
& \textit{Schwinger model parameters}\\
\colrule
$\theta$ & Lattice chiral transformation angle\\
$m$ & Lattice fermionic mass\\
$\mu_5$ & Lattice chiral chemical potential\\
$\omega$ & Half of inverse lattice spacing\\
\colrule
& \textit{Numerical results}\\
\colrule
$J$ & Electric current on the lattice\\
$J_q$ & Coupling term of $\sigma_q$ in $J$\\
$\ev{J}_\text{ed}$ & ED results of $J$ (10 Trotter steps) \\
$\ev{J}_\text{ED}$ & ED results of $J$ (100 Trotter steps)\\
$\ev{J}_\text{Noisy}$ & Noisy results of $J$\\
$\ev{J}_\text{MH}$ & Mitigation result of $J$\\
$L_\text{Trotter}$ & Trotter error bound of $J$ simulations\\
$L^r_\text{Noisy/MH}$ & Noisy/MH $J$ total error over ED\\
$P^r_\text{Noisy/MH}$ & Noisy/MH $J$ fit similarity over ED\\
$\va{p}_\text{ED/Noisy/MH}$ & Polynomial $J$ fit coefficients\\
$\eta$ & Noise level of $J$ simulations\\
\colrule
\end{longtable}

\section{Computation of the BBGKY equations}\label{app:commutator}
Start by denoting $\comm{A}{B}_{\pm} := AB \pm BA$ for any quantum operators $A$ and $B$ where, for generality purposes, we also consider the anticommutator case throughout the text. Remember also that same-site Pauli operators satisfy the product relation \cite{arfken2013mathematical}
\begin{equation}
    \sigma_k^\mu \sigma_k^\nu = \delta_{\mu\nu} + i \eps_{\mu\nu\lambda} \sigma_k^\lambda,
\end{equation}
where Einstein's summation convention applies to $\lambda \in \qty{1,2,3}$. In particular, this implies that same-site Pauli operators of different directions $\mu \neq \nu$ anticommute $\sigma_k^\mu \sigma_k^\nu = - \sigma_k^\nu \sigma_k^\mu$. Finally, define the number
\begin{equation}
    \delta_\pm := \begin{cases}
        0, & \text{if $+$}\\
        1, & \text{if $-$}
    \end{cases}.
\end{equation}
Ehrenfest's theorem states that \cite{Ehrenfest:1927swx}
\begin{equation}
    i \dv{t} \ev{\sigma^a_A} = \ev{\comm{\sigma^a_A}{H}_-}
\end{equation}
so, using \eqref{eq:hamiltonian_general} and the bilinearity of the commutator, one is interested in evaluating
\begin{equation}
\begin{split}
    &\comm{\sigma^a_A}{\sigma^b_B}_\pm=\\
    &= \qty(\prod_{i \in A \setminus B} \sigma^{a_i}_i) \comm{\prod_{i \in A \cap B} \sigma^{a_i}_i}{\prod_{j \in A \cap B} \sigma^{b_j}_j}_\pm \qty(\prod_{j \in B \setminus A} \sigma^{b_j}_j)\\
    &=\qty(\prod_{i \in A \setminus B} \sigma^{a_i}_i) \qty(\prod_{j \in B \setminus A} \sigma^{b_j}_j)\\
    &\quad\cdot \qty(\prod_{k \in A \cap B} \sigma^{a_k}_k \sigma^{b_k}_k \pm \prod_{k \in A \cap B} \sigma^{b_k}_k \sigma^{a_k}_k)\\
    &= \qty(1 \pm (-1)^{d^{ab}_{AB}})\\
    &\quad\cdot\qty(\prod_{i \in A \setminus B} \sigma^{a_i}_i) \qty(\prod_{j \in B \setminus A} \sigma^{b_j}_j)\qty(\prod_{k \in A \cap B} \sigma^{a_k}_k \sigma^{b_k}_k)\\
    &= i^{d^{ab}_{AB}} \qty(1 \pm (-1)^{d^{ab}_{AB}})\\
    &\quad\cdot\qty(\prod_{i \in A \setminus B} \sigma^{a_i}_i) \qty(\prod_{j \in B \setminus A} \sigma^{b_j}_j) \prod_{k \in A \cap B} \qty(\delta_{a_k b_k} + \eps_{a_k b_k c} \sigma_k^c)\\
    &= i^{\delta_\pm} f^{\pm ab}_{AB}\\
    &\quad\cdot\qty(\prod_{i \in A \setminus B} \sigma^{a_i}_i) \qty(\prod_{j \in B \setminus A} \sigma^{b_j}_j) \prod_{k \in A \cap B} \qty(\delta_{a_k b_k} + \eps_{a_k b_k c} \sigma_k^c).
\end{split}
\end{equation}
In the first equality, we partition $A \cup B = (A \setminus B) \cup (B \setminus A) \cup (A \cap B)$ and extract the $A \setminus B$ and $B \setminus A$ portions of the Pauli strings from the commutator. In the third equality, the anticommutation of same-site Paulis picks up a $(-1)$ factor if their directions differ, leading to the definition of the number of different directions $d^{ab}_{AB}$ in \eqref{eq:differences_number}. In the fourth equality, all imaginary units $i$ are extracted because the sum in the last product is exclusive, meaning that for every element of $A \cap B$ either the Kronecker delta or the Levi-Civita symbol part is picked. Finally, in the last equality, we compute
\begin{equation}
\begin{split}
    &i^{d^{ab}_{AB}} \qty(1 \pm (-1)^{d^{ab}_{AB}})=\\
    &= i^{\delta_\pm} i^{d^{ab}_{AB} - \delta_\pm} \cdot \begin{cases}
        2, & \text{if $(d_{AB}^{ab} - \delta_\pm) \in 2\mathbb{N}_0$}\\
        0, & \text{otherwise}
    \end{cases}\\
    &= i^{\delta_\pm} \underbrace{2(-1)^{(d_{AB}^{ab} - \delta_\pm)/2} \cdot \begin{cases}
        1, & \text{if $(d_{AB}^{ab} - \delta_\pm) \in 2\mathbb{N}_0$}\\
        0, & \text{otherwise}
    \end{cases}}_{=: f^{\pm ab}_{AB}},
\end{split}
\end{equation}
leading to the definition of the generalized factor $f^{\pm ab}_{AB}$, so that $f_{AB}^{ab} := f_{AB}^{- ab}$ in \eqref{eq:f_factor_def}. The generalized factor fully controls the result of the commutation/anticommutation
\begin{equation}
    f_{AB}^{\pm ab} = 0 \quad \Leftrightarrow \quad \comm{\sigma^a_A}{\sigma^b_B}_\pm = 0.
\end{equation}
Hence, for the $\delta_- = 1$ commutation choice, to obtain $\comm{\sigma^a_A}{\sigma^b_B}_- \neq 0$ it must be $d_{AB}^{ab} \geq 1$, implying $A \cap B \neq \varnothing$. In such cases, the $i^{\delta_-} = i$ factor cancels the imaginary unit present in Ehrenfest's theorem, in accordance with the fact that expectation values are real-valued quantities.

\section{Connections of the BBGKY hierarchy}\label{app:symmetric_hierarchy}
Let $\sigma_A^a$ and $\sigma_C^c$ be two Pauli strings. We define
\begin{equation}
\begin{split}
    &\sigma^a_A \to \sigma^c_C \quad \xLeftrightarrow{\text{def}}\\
    &\exists ((B,b), \gamma) \in \mathcal{H} \times (\mathbb{C} \setminus \qty{0}) \colon \comm{\sigma_A^a}{\sigma_B^b}_\pm = \gamma \sigma_C^c.
\end{split}
\end{equation}
It is always the case that $\comm{\sigma_A^a}{\sigma_B^b}_\pm$ is proportional to a single Pauli string because of the previous Appendix \ref{app:commutator} computation. Moreover, the proportionality constant $\gamma \neq 0$ simultaneously ensures that $f^{\pm ab}_{AB} \neq 0$ and that there exists a $t \in [0, T]$ such that $h_B^b(t) \neq 0$.

We now show that $\sigma^a_A \to \sigma^c_C$ implies $\sigma^c_C \to \sigma^a_A$. Assuming $\sigma^a_A \to \sigma^c_C$, multiply $\comm{\sigma_A^a}{\sigma_B^b}_\pm = \gamma \sigma_C^c$ respectively from the left and from the right by the (fixed) $\sigma_B^b$ to, respectively, obtain
\begin{equation}
\begin{split}
    \sigma_B^b \sigma_A^a \sigma_B^b \pm \sigma_A^a &= \gamma \sigma_B^b \sigma_C^c,\\
    \sigma_A^a \pm \sigma_B^b \sigma_A^a \sigma_B^b &= \gamma \sigma_C^c \sigma_B^b,
\end{split}
\end{equation}
where we used $(\sigma_A^a)^2 = (\sigma_B^b)^2 = 1$. Combining the above two equations gives
\begin{equation}
    \gamma \comm{\sigma_C^c}{\sigma_B^b}_\pm = 2\sigma_A^a \pm 2 \sigma_B^b \sigma_A^a \sigma_B^b.
\end{equation}
We now show that the above is not null. Assume by contradiction that $\sigma_A^a \pm  \sigma_B^b \sigma_A^a \sigma_B^b = 0$, then by multiplying both sides from the right by $\sigma_B^b$ we obtain $\comm{\sigma_A^a}{\sigma_B^b}_\pm = 0$, in contradiction with the initial $\gamma \neq 0$ assumption. By the previous Appendix \ref{app:commutator} computation, $\gamma \comm{\sigma_C^c}{\sigma_B^b}_\pm \neq 0$ must be proportional to a single Pauli string as well. Finally, since $\sigma_A^a \neq \sigma_B^b \sigma_A^a \sigma_B^b$, then $\sigma_B^b \sigma_A^a \sigma_B^b$ must be proportional to $\sigma_A^a$ by a non-unit proportionality constant, leading to $\comm{\sigma_C^c}{\sigma_B^b}_\pm$ being proportional to $\sigma_A^a$, which is the desired result.

\section{Bounds on correlator lengths}\label{app:bounds}
Let $L^{\pm ab}_{AB}$ denote the length of the $\bigl \langle \comm{\sigma^a_A}{\sigma^b_B}_\pm \bigr \rangle$ correlator, meaning $L^{ab}_{AB} := L^{- ab}_{AB}$ in Sec.~\ref{sec:hierarchical_structure}. We assume $f_{AB}^{\pm ab} \neq 0$ so, from Appendix \ref{app:commutator}, in the $\delta_- = 1$ case it must be
$A \cap B \neq \varnothing$. This leads to the lower bound
\begin{equation}\label{eq:appendix_inequality}
    \delta_\pm \leq \abs{A \cap B} \leq \min(\abs{A}, \abs{B}),
\end{equation}
the upper bound being a trivial inequality.

If all directions in $A \cap B$ differ, then $\comm{\sigma^a_A}{\sigma^b_B}_\pm$ is composed of at most $\abs{A \cup B}$ Pauli operators. Likewise, if all directions in $A \cap B$ match, then $\comm{\sigma^a_A}{\sigma^b_B}_\pm$ is composed of at least $\abs{A \Delta B} + \delta_\pm$ Pauli operators, where the addition of $\delta_\pm$ stems directly from the $d^{ab}_{AB} \geq 1$ requirement in the $\delta_- = 1$ case, and where $A \Delta B := (A \setminus B) \cup (B \setminus A) = (A \cup B) \setminus (A \cap B)$ is the symmetric set difference. Therefore, $L^{\pm ab}_{AB}$ is bounded by
\begin{equation}
    \abs{A} + \abs{B} - 2\abs{A \cap B} + \delta_\pm \leq L^{\pm ab}_{AB} \leq \abs{A} + \abs{B} - \abs{A \cap B}.
\end{equation}
By using \eqref{eq:appendix_inequality} and the identity $\abs{\abs{A} - \abs{B}} = \abs{A} + \abs{B} - 2\min(\abs{A}, \abs{B})$ \cite{spivak2006calculus}, the tightest bounds for $L^{\pm ab}_{AB}$ become
\begin{equation}\label{eq:appendix_bounds}
    \abs{\abs{A} - \abs{B}} + \delta_\pm \leq L^{\pm ab}_{AB} \leq \abs{A} + \abs{B} - \delta_\pm.
\end{equation}
As a check, notice how the lower bound is always non-negative. Also, we check that in both $\delta_\pm \in \qty{0,1}$ cases the upper bound is always bigger than or equal to the lower one. Indeed, in the $\delta_+ = 0$ case we recover the triangle inequality, while in the $\delta_- = 1$ case either $\abs{A} \geq \abs{B}$, implying $\abs{B} \geq 1$, or $\abs{B} \geq \abs{A}$, implying $\abs{A} \geq 1$.

\section{Derivation of the BBGKY action}\label{app:action_derivation}
The equilibrium configuration $\va{\chi}$ is such that $\forall (A,a) \in \mathcal{Q}_R: \forall s\in\qty{0,\dots,N_\text{T}}: E_{A}^{a}(s, \va{\chi}) = 0$, with fixed boundary conditions $\forall q \in \qty{1, \dots, \abs{\mathcal{Q}_R}}: \chi_{q0} = \bar{x}_{q0}$. Then, for any $q$ and $s > 0$, it is
\begin{equation}
\begin{split}
    &\chi_{qs} = \int \dd{\va{x}} x_{qs} \prod_{(A, a) \in \mathcal{Q}_R} \qty(\prod_{s'=0}^{N_\text{T}} \delta(E^{a}_A (s', \va{x})))\\
    &= \qty(\prod_{s'=0}^{N_\text{T}} \lim_{\lambda_{Aa}^{s'} \to \infty}) \int \dd{\va{x}} x_{qs} \prod_{\substack{(A, a) \in \mathcal{Q}_R\\1 \leq s' \leq N_\text{T}}} \sqrt{\frac{\lambda_{Aa}^{s'}}{\pi}} e^{-\lambda_{Aa}^{s'} E_A^{a}(s',\va{x})^2}\\
    &= \lim_{\lambda \to \infty} \sqrt{\frac{\lambda T}{\pi N_\text{T}}}^{\abs{\mathcal{Q}_R} (N_\text{T} + 1)}\\
    &\qquad\quad\cdot\int \dd{\va{x}} x_{qs} \exp(-\lambda \Delta t \sum_{(A, a) \in \mathcal{Q}_R} \sum_{s'=0}^{N_\text{T}} E_A^{a}(s', \va{x})^2)\\
    &= \lim_{\lambda \to \infty} \int \dd{\va{x}} (\va{x})_{qs} w(\lambda) e^{-\lambda S(\va{x})}.
\end{split}
\end{equation}
In the second equality, we represent the Dirac delta distributions with \cite{arfken2013mathematical}
\begin{equation}
    \delta_\lambda(x) = \sqrt{\frac{\lambda}{\pi}} e^{-\lambda x^2},
\end{equation}
where $\lambda$ is the parameter controlling the faithfulness of the delta function representation $\lim_{\lambda \to \infty} \delta_\lambda(x) = \delta(x)$. As there is a Dirac delta for every equation $(A,a)$ at every time point $s'$, each Dirac delta representation is controlled by its own $\lambda^{s'}_{Aa}$ parameter. In the third equality, we rescale and collect all these parameters into a unique global $\lambda$. In the fourth equality, we define the factor $w(\lambda) := \sqrt{\lambda T/(\pi N_\text{T})}^{\abs{\mathcal{Q}_R} (N_\text{T} + 1)}$, the BBGKY action \eqref{eq:bbgky_action}, and set $S(\va{x}) = S_\text{B}(\va{x})$. Finally, we define the \eqref{eq:montecarlo_weight} weight $\omega(\lambda, \va{x}) := w(\lambda) e^{-\lambda S(\va{x})}$. As a last remark, notice how the $\chi_{qs}$ are indirectly a function of the fixed initial conditions $\bar{x}_{qs}$, as they appear in all $E^a_A(s,\va{x})$ with $s=0$ or $s=1$ via the approximation of the time derivative \eqref{eq:discrete_derivative}.

\section{Variance of Z-measurements}\label{app:z_measurements}

Let $B_1, \dots, B_{N_\text{S}} \overset{\text{iid}}{\sim} \mathcal{B}(p)$ be $N_\text{S}$ discrete Bernoulli random variables of identical parameter $p \in [0, 1]$. Then, for any $1 \leq i \leq N_\text{S}$, the discrete $X_i := 2B_i - 1$ random variable represents a Z-measurement (or one-shot measurement) of a quantum observable, whose explicit definition fixes the value of $p$. The expectation value $\mathbb{E}[X_i] = 2p-1 \in [-1,1]$ of $X_i$ corresponds to the expectation value of the quantum observable, and one can find that
\begin{equation}
    \text{Var}[X_i] = 4p(1-p) = 1 - \mathbb{E}[X_i]^2 \in [0,1].
\end{equation}

Given the realization $\bar{x}_{qs}$ of $\mathbb{E}[X_i]$, we pick the realization $y^2_{qs} = 1 - \bar{x}^2_{qs}$ of $\text{Var}[X_i]$ as the standard deviation $y_{qs}$ of the Gaussian penalty term in \eqref{eq:quantum_action} because we want $x_{qs}$ to be a realization of a normally distributed $\mathcal{N}(\bar{x}_{qs}, y^2_{qs})$ continuous random variable. Indeed, we do not want $x_{qs}$ to be a realization of the average $A := \sum^{N_\text{S}}_{j=1} X_j/N_\text{S}$ discrete random variable, because in that case $\text{Var}[A] = \text{Var}[X_i]/N_\text{S}$, and as a result $y_{qs}$ would tend to zero in the $N_\text{S} \to \infty$ limit. This allows $x_{qs}$ to probe a fixed range around the noisy $\bar{x}_{qs}$, whose quantum error is independent of shot noise.

Divisions by zero in \eqref{eq:quantum_action} can still occur if $\bar{x}_{qs} \in \qty{\pm 1}$, leading to $y_{qs} = 0$, which can happen with an unlikely (but still possible) sequence of identical Z-measurements. To avoid that, in the $\bar{x}_{qs} \in \qty{\pm 1}$ cases we slightly modify the estimation of the average used in \eqref{eq:variance}, so that the former is forced to include an additional different Z-measurement
\begin{equation}
    \bar{x}_{qs} \to \frac{N_\text{S} \bar{x}_{qs} - \text{sgn}(\bar{x}_{qs})}{N_\text{S} + 1}.
\end{equation}
One readily checks that the above expression tends to $\bar{x}_{qs}$ in the $N_\text{S} \to \infty$ limit.

\section{Intensity of the noise model}\label{app:intensity_noise}

\begin{figure}
\centering
\includegraphics[width=\linewidth]{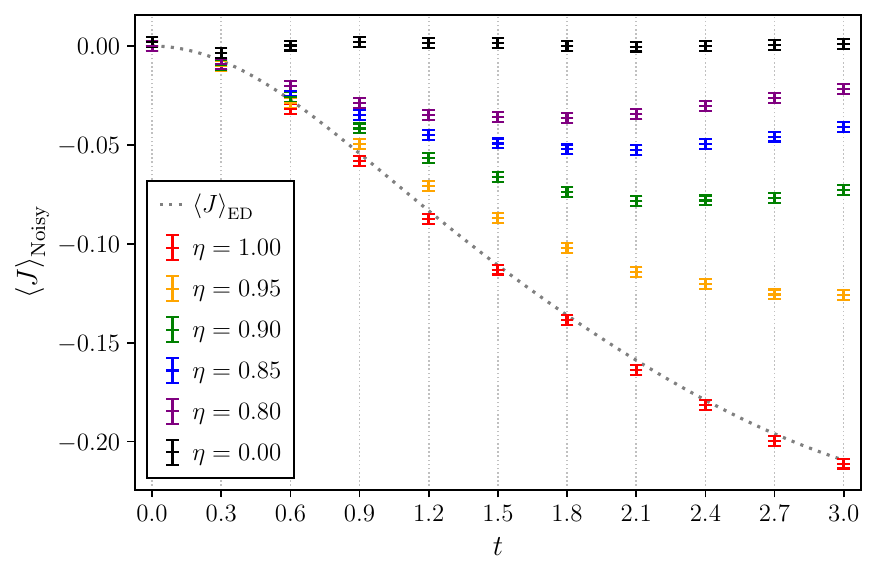}
\caption{Effect of \eqref{eq:noisemodel} on $\ev{J}_\text{Noisy}$ measurements. Different choices of $\eta$ lead to different noise models, spanning the spectrum of quantum error from noiseless QPUs ($\eta = 1$) to current NISQ noise ($\eta = 0$). We emphasize that the $\eta = 0.9$ results are the same as the $(m, \mu_5, r) = (0.5, 0.2, 2)$ simulation in Fig.~\ref{fig:resultsb}.}
\label{fig:superimposed_errors}
\end{figure}

Let $\bar{x}_{qs}$ be the quantum measurement of $\ev{\sigma_q(t_s)}$, obtained from a simulated QPU under the influence of a realistic noise model. Denote by $\Tilde{x}_{qs}$ the noiseless realization of $\bar{x}_{qs}$, that is, the quantum measurement of $\ev{\sigma_q(t_s)}$ obtained from a noiseless simulation of the QPU. $\Tilde{x}_{qs}$ is affected by Trotter error and shot noise. To mimic the suppression of the realistic noise model, at every measurement we systematically apply the substitution
\begin{equation}\label{eq:noisemodel}
    \bar{x}_{qs} \to (1-\eta^s) \bar{x}_{qs} + \eta^s \Tilde{x}_{qs},
\end{equation}
where $\eta \in [0, 1]$ characterizes the desired noiselessness of the noise model. This is because, as expected from the generally observed accumulation of quantum noise, we want the importance of the noisy contribution to grow with $s$, regardless of the value of $\eta$. The above can effectively be considered as a noise model of its own, and in Fig.~\ref{fig:superimposed_errors} we showcase its effect on measurements. All quantum simulation results in this work are obtained with $\eta = 0.9$.


\bibliography{bio}

\end{document}